\documentclass[a4paper,11pt]{article}
\usepackage[left=3.17cm,right=3.17cm,top=2.54cm,
headheight=0.5cm,headsep=0.54cm,bottom=2.54cm,footskip=0.79cm
]{geometry}

\usepackage[all]{xy}
\usepackage{amsmath,amsfonts,amssymb,amsthm,epsfig,amscd,comment,latexsym,psfrag}
\usepackage{nicefrac,xspace,tikz}

\usetikzlibrary{arrows}
\usetikzlibrary{decorations.markings}
\def\Z{\mathbb Z}

\def\C{\mathbb C}

\def\1{{\bf{1}}}

\usepackage[pdfstartview=FitH]{hyperref}

\def\footnoterule{\kern 1mm \hrule width 7cm \kern 2.2mm}%

\newcommand{\bea}{\begin{eqnarray}}
\newcommand{\eea}{\end{eqnarray}}
\newcommand{\beaa}{\begin{eqnarray*}}
\newcommand{\eeaa}{\end{eqnarray*}}
\newcommand{\be}{\begin{equation}}
\newcommand{\ee}{\end{equation}}
\newcommand{\nn}{\nonumber}

\def\3Dtwoboxy{\begin{tikzpicture}
\draw (0,0)rectangle(0.5,0.25);
\draw (0.25,0) -- (0.25,0.25) [-];
\draw (0,0.25) -- (0.175,0.35) [-];
\draw [shift = {+(0.25,0)}](0,0.25) -- (0.175,0.35) [-];
\draw [shift = {+(0.5,0)}](0,0.25) -- (0.175,0.35) [-];
\draw [shift = {+(0.5,-0.250)}](0,0.25) -- (0.175,0.35) [-];
\draw (0.175,0.35) -- (0.675,0.35) [-];
\draw [shift = {+(0.25,0)}] (0.425,0.35) -- (0.425,0.1) [-];
\end{tikzpicture}}

\begin{document}

\title{KP hierarchy, affine Yangian and $W_{1+\infty}$ algebra}
\author{Wang Na\footnote{Corresponding author: wangna@henu.edu.cn }\\
\small School of Mathematics and Statistics, Henan University, Kaifeng, 475001, China}

\date{}
\maketitle

\begin{abstract}
3D (dimensional) Young diagrams are a generalization of 2D Young diagrams. We want to construct the structures on 3D Young diagrams paralleled to that on 2D Young diagrams. We have already obtained the 3D Bosons, 3D Fermions and 3-Jack polynomials, which are the symmetric functions on 3D Young diagrams. In this paper, we realize the KP hierarchy by the fields $V_j(z)$ in the $W_{1+\infty}$ algebra, where $V_j(z)$ are constructed by $J_j(z)$, which are the Bosonic fields associated the 2D Young diagrams on the slices $z=j$ of 3D Young diagrams. We also show the correspondence between the Bosonic Fock spaces and Fermionic Fock space for the cases: Schur functions, Jack polynomials and 3-Jack polynomials.
\end{abstract}

\section{Introduction}\label{sect0}
Schur functions are a famous kind of symmetric functions defined on 2D Young diagrams. Let the power sum $p_n=\sum_i t_i^n=nx_n$ and
${\bf x}=(x_1,x_2,\cdots)$.
The operators $S_n({\bf x})$ are determined by the generating function:
\be\label{hxi}
e^{\xi({\bf x},k)}:=\sum_{n=0}^\infty S_n({\bf x})k^n,\quad\text{where}\ \  \xi({\bf x},k)=\sum_{n=1}^\infty x_n k^n
\ee
and set $S_n({\bf x})=0$ for $n<0$.
Note that $S_n({\bf t})$ is the complete homogeneous symmetric function \begin{equation}\label{compsym}
\sum _{{ i_{1}\leq i_{2}\leq \cdots \leq i_{n}}}t_{{i_{1}}}t_{{i_{2}}}\cdots t_{{i_{n}}}.
\end{equation}
For Young diagrams $\lambda=(\lambda_1,\lambda_2,\cdots,\lambda_l)$, the Schur function $S_{\lambda}=S_{\lambda}({\bf x})$ is a polynomial in $\C[{\bf x}]$ defined by the Jacobi-Trudi formula \cite{Mac}:
\be\label{Slambda}
S_{\lambda}({\bf x})=\text{det}\left(
h_{\lambda_{i}-i+j}({\bf x})
 \right)_{1\leq i,j\leq l}.
\ee
Schur functions and 2D Young diagrams have played a prominent role in mathematics (representation theory, combinatorics, especially in integrable models) for a long time\cite{Stan,FH,MJD}.

The KP hierarchy \cite{DKJM} is one of the most important integrable hierarchies and it arises in many different fields of mathematics and physics such as enumerative algebraic geometry, topological field and string theory. Recently, there are many generalizations of the KP hierarchy\cite{JR, Tsuda,Tsuda1}. Kyoto school use Schur functions in a remarkable way to understand the KP and KdV hierarchies\cite{MJD}.
It is known that the KP hierarchy can be realized by the fields $V_j(z)$ in the $W_{1+\infty}$ algebra which are constructed by one Bosonic field $J(z)$ \cite{Minru}. In this paper, we generalize this realization of KP hierarchy to the cases of fields $V_j(z)$ which are symmetric about coordinate axes.

This symmetry has appeared in Schur functions, that is, the expressions of Schur functions $S_\lambda(x)$ and $S_{\lambda'}(x)$ are similar, where Young diagram $\lambda'$ is the conjugate of $\lambda$. For example,
  \beaa
&&S_{\begin{tikzpicture}
\draw [step=0.2](0,0) grid(.6,.2);
\draw [step=0.2](0,-0.2) grid(.2,0);
\end{tikzpicture}}= \frac{1}{8}(p_1^4+2p_1^2p_{2}-p_{2}^2-2p_{4}),\\
&&S_{\begin{tikzpicture}
\draw [step=0.2](0,0) grid(.4,.2);
\draw [step=0.2](0,-0.2) grid(.2,0);
\draw [step=0.2](0,-0.4) grid(.2,-0.2);
\end{tikzpicture}}=\frac{1}{8}(p_1^4-2p_1^2p_{2}-p_{2}^2+2p_{4}).
\eeaa
 Then we believe that the Schur functions should satisfy some symmetry about the coordinate axes $x$-axis and $y$-axis. We associate $h_1$ and $h_2$ to $y$-axis and $x$-axis respectively. We have defined the symmetric functions $Y_\lambda$ \cite{WBCW} which are symmetric about $x$-axis and $y$-axis (that is, they are symmetric about $h_1$ and $h_2$), and become the Jack polynomials when $h_1=\sqrt{\alpha},h_2=-1/\sqrt{\alpha}$. In this paper, we construct the fields $V_j(z)$ in the $W_{1+\infty}$ algebra, which are symmetric about $x$-axis and $y$-axis, and realize the KP hierarchy using the fields $V_j(z)$.

3D Young diagram is a generalization of 2D Young diagram, which arose naturally in crystal melting model\cite{ORV,NT}. 3D Young diagrams also have many applications in many fields of mathematics and physics, such as statistical models, number theory, representations of some algebras (Ding-Iohara-Miki algebras, affine Yangian, etc). Every slice of a 3D Young diagram is a 2D Young diagram, or we treat a 2D Young diagram as a special 3D Young diagram which has one layer. We consider 3D Young diagrams which have at most $N$ layers in $z$-axis direction. Since we have already constructed 3-Jack polynomials which are symmetric functions defined on 3D Young diagrams. In this paper, we construct the field $V_4(z)$ in the $W_{1+\infty}$ algebra. We find that the OPEs $V_j(z)V_k(w)$ here are symmetric about $x$-axis, $y$-axis and $z$-axis, and when $N=1$, the the fields $V_j(z)$ here become $V_j(z)$ which are symmetric about $x$-axis and $y$-axis. We also realize the KP hierarchy by $V_j(z)$ which are symmetric about three coordinate axes.

 The paper is organized as follows. In section \ref{sect1}, we recall the definition of affine Yangian of ${\mathfrak{gl}}(1)$ and its representation on 3D Young diagrams. In section \ref{sect2}, we recall the construction of $W_{1+\infty}$ algebra from the Miura transformation, and we construct the field $V_4(z)$.  In section \ref{sect3},  we realize the KP hierarchy from the special case $N=1, h_1=1,h_2=-1$ of the fields in the $W_{1+\infty}$ algebra constructed from the Miura transformation. In section \ref{sect4}, we realize the KP hierarchy from the special case $N=1$ of the fields in the $W_{1+\infty}$ algebra constructed from the Miura transformation. In section \ref{sect5}, we realize the KP hierarchy by the general fields in the $W_{1+\infty}$ algebra constructed from the Miura transformation. In section \ref{sect6}, we give the correspondence between the Fermionic Fock space and the Bosonic Fock space for three cases: the Schur functions, the Jack polynomials, and the 3-Jack polynomials.
\section{Affine Yangian of ${\mathfrak{gl}}(1)$}\label{sect1}
In this section, we recall the affine Yangian of ${\mathfrak{gl}}(1)$ and its representation on the space of 3D Young diagrams. Let $h_1,h_2$ and $h_3$ be three complex numbers satisfying $h_1+h_2+h_3=0$. Define
\beaa
\sigma_2 &=& h_1 h_2 + h_1 h_3 + h_2 h_3,\\
 \sigma_3 &=& h_1 h_2 h_3.
\eeaa
We associate $h_1,\ h_2,\ h_3$ to $y,\ x,\ z$-axis respectively.

The affine Yangian $\mathcal{Y}$ of ${\mathfrak{gl}}(1)$ is an associative algebra with generators $e_j, f_j$ and $\psi_j$, $j = 0, 1, \ldots$ and the following relations\cite{Pro,Tsy}
\begin{eqnarray}
&&\left[ \psi_j, \psi_k \right] = 0,\\
&&\left[ e_{j+3}, e_k \right] - 3 \left[ e_{j+2}, e_{k+1} \right] + 3\left[ e_{j+1}, e_{k+2} \right] - \left[ e_j, e_{k+3} \right]\nonumber \\
&& \quad + \sigma_2 \left[ e_{j+1}, e_k \right] - \sigma_2 \left[ e_j, e_{k+1} \right] - \sigma_3 \left\{ e_j, e_k \right\} =0,\label{yangian1}\\
&&\left[ f_{j+3}, f_k \right] - 3 \left[ f_{j+2}, f_{k+1} \right] + 3\left[ f_{j+1}, f_{k+2} \right] - \left[ f_j, f_{k+3} \right] \nonumber\\
&& \quad + \sigma_2 \left[ f_{j+1}, f_k \right] - \sigma_2 \left[ f_j, f_{k+1} \right] + \sigma_3 \left\{ f_j, f_k \right\} =0, \label{yangian2}\\
&&\left[ e_j, f_k \right] = \psi_{j+k},\label{yangian3}\\
&& \left[ \psi_{j+3}, e_k \right] - 3 \left[ \psi_{j+2}, e_{k+1} \right] + 3\left[ \psi_{j+1}, e_{k+2} \right] - \left[ \psi_j, e_{k+3} \right]\nonumber \\
&& \quad + \sigma_2 \left[ \psi_{j+1}, e_k \right] - \sigma_2 \left[ \psi_j, e_{k+1} \right] - \sigma_3 \left\{ \psi_j, e_k \right\} =0,\label{yangian4}\\
&& \left[ \psi_{j+3}, f_k \right] - 3 \left[ \psi_{j+2}, f_{k+1} \right] + 3\left[ \psi_{j+1}, f_{k+2} \right] - \left[ \psi_j, f_{k+3} \right] \nonumber\\
&& \quad + \sigma_2 \left[ \psi_{j+1}, f_k \right] - \sigma_2 \left[ \psi_j, f_{k+1} \right] + \sigma_3 \left\{ \psi_j, f_k \right\} =0,\label{yangian5}
\end{eqnarray}
together with boundary conditions
\begin{eqnarray}
&&\left[ \psi_0, e_j \right]  = 0, \left[ \psi_1, e_j \right] = 0,  \left[ \psi_2, e_j \right]  = 2 e_j ,\label{yangian6}\\
&&\left[ \psi_0, f_j \right]  = 0,  \left[ \psi_1, f_j \right]  = 0,  \left[ \psi_2, f_j \right]  = -2f_j ,\label{yangian7}
\end{eqnarray}
and a generalization of Serre relations
\begin{eqnarray}
&&\mathrm{Sym}_{(j_1,j_2,j_3)} \left[ e_{j_1}, \left[ e_{j_2}, e_{j_3+1} \right] \right]  = 0, \label{yangian8} \\
&&\mathrm{Sym}_{(j_1,j_2,j_3)} \left[ f_{j_1}, \left[ f_{j_2}, f_{j_3+1} \right] \right]  = 0,\label{yangian9}
\end{eqnarray}
where $\mathrm{Sym}$ is the complete symmetrization over all indicated indices which include $6$ terms.

The affine Yangian $\mathcal{Y}$ has a representation on 3D Young diagrams.
As in our paper \cite{3DFermionYangian}, we use the following notations. For a 3D Young diagram $\pi$, the notation $\Box\in \pi^+$ means that this box is not in $\pi$ and can be added to $\pi$. Here ``can be added'' means that when this box is added, it is still a 3D Young diagram. The notation $\Box\in \pi^-$ means that this box is in $\pi$ and can be removed from $\pi$. Here ``can be removed" means that when this box is removed, it is still a 3D Young diagram. For a box $\Box$, we let
\begin{equation}\label{epsilonbox}
h_\Box=h_1y_\Box+h_2x_\Box+h_3z_\Box,
\end{equation}
where $(x_\Box,y_\Box,z_\Box)$ is the coordinate of box $\Box$ in coordinate system $O-xyz$. Here we use the order $y_\Box,x_\Box,z_\Box$ to match that in paper \cite{Pro}.

Following \cite{Pro,Tsy}, we introduce the generating functions:
\begin{eqnarray}
e(u)&=&\sum_{j=0}^{\infty} \frac{e_j}{u^{j+1}},\nonumber\\
f(u)&=&\sum_{j=0}^{\infty} \frac{f_j}{u^{j+1}},\\
\psi(u)&=& 1 + \sigma_3 \sum_{j=0}^{\infty} \frac{\psi_j}{u^{j+1}},\nonumber
\end{eqnarray}
where $u$ is a parameter.
Introduce
\begin{equation}\label{psi0}
\psi_0(u)=\frac{u+\sigma_3\psi_0}{u}
\end{equation}
and
\begin{eqnarray} \label{dfnvarphi}
\varphi(u)=\frac{(u+h_1)(u+h_2)(u+h_3)}{(u-h_1)(u-h_2)(u-h_3)}.
\end{eqnarray}
For a 3D Young diagram $\pi$, define $\psi_\pi(u)$ by
\begin{eqnarray}\label{psipiu}
\psi_\pi(u)=\psi_0(u)\prod_{\Box\in\pi} \varphi(u-h_\Box).
\end{eqnarray}
In the following, we recall the representation of the affine Yangian on 3D Young diagrams as in paper \cite{Pro} by making a slight change. The representation of affine Yangian on 3D Young diagrams is given by
\begin{eqnarray}
\psi(u)|\pi\rangle&=&\psi_\pi(u)|\pi\rangle,\\
e(u)|\pi\rangle&=&\sum_{\Box\in \pi^+}\frac{E(\pi\rightarrow\pi+\Box)}{u-h_\Box}|\pi+\Box\rangle,\label{eupi}\\
f(u)|\pi\rangle&=&\sum_{\Box\in \pi^-}\frac{F(\pi\rightarrow\pi-\Box)}{u-h_\Box}|\pi-\Box\rangle\label{fupi}
\end{eqnarray}
where $|\pi\rangle$ means the state characterized by the 3D Young diagram $\pi$ and the coefficients
\begin{equation}\label{efpi}
E(\pi \rightarrow \pi+\square)=-F(\pi+\square \rightarrow \pi)=\sqrt{\frac{1}{\sigma_3} \operatorname{res}_{u \rightarrow h_{\square}} \psi_\pi(u)}
\end{equation}
 Equations (\ref{eupi}) and (\ref{fupi}) mean generators $e_j,\ f_j$ acting on the 3D Young diagram $\pi$ by
\begin{equation}
\begin{aligned}
e_j|\pi\rangle &=\sum_{\square \in \pi^{+}} h_{\square}^j E(\pi \rightarrow \pi+\square)|\pi+\square\rangle,
\end{aligned}
\end{equation}
\begin{equation}
\begin{aligned}
f_j|\pi\rangle &=\sum h_{\square}^j F(\pi \rightarrow \pi-\square)|\pi-\square\rangle .
\end{aligned}
\end{equation}
In the following of this paper, we treat $E(\pi \rightarrow \pi+\square)|\pi+\square\rangle$ as one element and still denote it by $|\pi+\square\rangle$, then 3D Young diagrams depend on the box growth process.
\section{Miura transformation and $W_{1+\infty}$ algebra}\label{sect2}
We begin this section by the definition of $W_{1+\infty}$ algebra. The $W_{1+\infty}$ algebra contains the Heisenberg algebra, the Virasoro algebra as  subalgebras\cite{Pro}. The generators are $V_{j,m}$ for $j\in\Z_+$, $m\in\Z$. The relations are
\bea
[V_{1,m}, V_{1,n}]&=&m\delta_{m+n,0}c_{1},\eea
\bea
[V_{2,m}, V_{2,n}]&=&(m-n) V_{2,m+n}+\frac{m^3-m}{12} \delta_{m+n,0} c_{2},
\eea
\bea
[V_{2,m}, V_{1,n}]&=&-n V_{1,m+n},
\eea
and generally
\be\label{vjmvkn}
[V_{j,m}, V_{k,n}]=\sum\limits_{\substack{0\leq l\leq j+k-2 \\ j+k-l\text{even}}}C_{jk}^lN_{jk}^l(m,n)V_{l,m+n},
\ee
where the coefficients $N_{jk}^l(m,n)$ are
\bea
N_{jk}^0(m,n)&=&\left(\begin{array}{cc}m+j-1\\j+k-1\end{array}\right)\delta_{m+n,0},\\
N_{jk}^l(m,n)&=&\sum_{s=0}^{j+k-l-1}\frac{(-1)^s}{(j+k-l-1)!(2l)_{j+k-l-1}}\left(\begin{array}{cc}j+k-l-1\\s\end{array}\right)\times\nn\\
&&[j+m-1]_{j+k-l-1-s}[j-m-1]_{s}[k+n-1]_{s}[k-n-1]_{j+k-l-1-s},\nn
\eea
and the structure constants $C_{jk}^l$ are
\bea
C_{jk}^0&=&\frac{(j-1)!^2(2j-1)!}{4^{j-1}(2j-1)!!(2j-3)!!}\delta_{jk}c_{j},\\
C_{jk}^l&=&\frac{1}{2\times 4^{j+k-l-2}}(2l)_{j+k-l-1}\times {}_4F_3\left(\begin{array}{cc}\frac{1}{2},\frac{1}{2},-\frac{1}{2}(j+k-l-2),-\frac{1}{2}(j+k-l-1)\\ \frac{3}{2}-j,\frac{3}{2}-k,\frac{1}{2}+l\end{array};1\right),\nn
\eea
with
\bea
(a)_n&=&a(a+1)\cdots (a+n-1),\\
{[a]}_n&=&a(a-1)\cdots (a-n+1),\\
{}_mF_n\left(\begin{array}{cc}a_1,\cdots,a_m\\ b_1,\cdots,b_n\end{array};z\right)&=&\sum_{k=0}^\infty\frac{(a_1)_k\cdots (a_m)_k}{(b_1)_k\cdots (b_n)_k}\frac{z^k}{k!}.
\eea
Note that here we allow that the central charges can be different.

Let
\be
B(z)=\sum_{n\in\Z}b_nz^{-n-1}
\ee
be the Bosonic field with the relation
\be
[b_n,b_m]=n\delta_{n+m,0}.
\ee
Define $V_j(z)=\sum_{n\in\Z}V_{j,n}z^{-j-n}$. It is known that the fields $V_{j}(z)$ have an representation by $B(z)$
\bea
{V}_1(z)&=&B(z),\nn\\
{V}_2(z)&=&\frac{1}{2}:B(z)^2:,\nn\\
{V}_3(z)&=&\frac{1}{3}:B(z)^3:,\label{V4N=1}\\
{V}_4(z)&=&\frac{1}{4}:B(z)^4:-\frac{3}{20}:B'(z)^2:+\frac{1}{10}:B''(z)B(z):.\nn
\eea
For Bosons $b_n$, the Bosonic Fock space is the space of 2D Young diagrams, or the space of Schur functions defined on 2D Young diagrams. Then the field $V_j(z)$ of the $W_{1+\infty}$ algebra can be represented on the space of 2D Young diagrams.

We consider $J_j(z)$
\be
J_j(z)=\sum_{n\in\Z}a_{j,n}z^{-n-1}
\ee
with the relation
\be\label{ajnakmcom}
[a_{j,n},a_{k,m}]=-\frac{1}{h_1h_2}\delta_{j,k}n\delta_{n+m,0}.
\ee

3D Young diagram can be treated as a series of 2D Young diagrams (here we consider the slices on plane $z=j$ of a 3D Young diagram, then every slice is a 2D Young diagram). Let $J_j(z)$ be the Bosonic field associated to the 2D Young diagrams on the slice $z=j$ of 3D Young diagrams. If we want that the fields $V_j(z)$ of the $W_{1+\infty}$ algebra have the representation on the space of 3D Young diagrams,
The fields $V_{j}(z)$ should be represented by a series of Bosonic fields $J_j(z), \ j=1,2,\cdots,N$, which corresponds to that a 3D Young diagram can be represented by a series of 2D Young diagrams. Here we allow the coefficients in $W_{1+\infty}$ algebra can be slightly different from that in the definition of $W_{1+\infty}$ algebra above.

Define
\be\label{j(z)}
J(z)=\sum_{n\in\Z}a_{n}z^{-n-1}=J_1(z)+J_2(z)+\cdots+J_N(z),
\ee
then the bosons $a_{n}$ satisfy
\be
[a_{n},a_{m}]=-\frac{N}{h_1h_2}n\delta_{n+m,0}=\psi_0n\delta_{n+m,0}.
\ee

Let $\alpha_0$ be a parameter,
 define the operator $U_k(z)$ as in \cite{Pro1} by the Miura transformation
\be
:(\alpha_0\partial+J_1(z))(\alpha_0\partial+J_2(z))\cdots (\alpha_0\partial+J_N(z)):=\sum_{k=0}^N U_k(z)(\alpha_0\partial)^{N-k}.
\ee
The fields $U_k(z)$ generate an algebra, which is $W_{1+N}$. The fields $V_n(z)$ can be realized by $U_k(z)$. We list the concrete expressions of the first few $U_k(z)$ as in \cite{pro1411}
\begin{eqnarray}
U_0 & = & 1, \\
U_1 & = & \sum_{j=1}^N J_j, \\
U_2 & = & \sum_{j<k} :J_j J_k : + \alpha_0 \sum_{j=1}^N (j-1) J_j^\prime, \\
U_3 & = & \sum_{j<k<l} :J_j J_k J_l: + \alpha_0 \sum_{j<k} (j-1) :J_j^\prime J_k: \\
\nonumber
& & + \alpha_0 \sum_{j<k} (k-2) :J_j J_k^\prime: + \frac{\alpha_0^2}{2} \sum_{j=1}^N (j-1)(j-2) J_j^{\prime\prime},\\
U_4 & = & \sum_{j<k<l<m} :J_j J_k J_l J_m: + \frac{\alpha_0^3}{6} \sum_j (j-1)(j-2)(j-3) J_j^{\prime\prime\prime} \\
\nonumber
& & + \alpha_0 \sum_{j<k<l} (j-1) :J_j^\prime J_k J_l: + (k-2) J_j J_k^\prime J_l: + (l-3) :J_j J_k J_l^\prime: \\
\nonumber
& & + \frac{\alpha_0^2}{2} \sum_{j<k} (j-1)(j-2) :J_j^{\prime\prime} J_k: + 2(j-1)(k-3) :J_j^\prime J_k^\prime: + (k-2)(k-3) :J_j J_k^{\prime\prime}.
\end{eqnarray}
Note that the expressions of $U_k(z)$ is the same with that in \cite{pro1411}, but the commutation relation of Boson fields $J_j(z)$ are slightly different from that in \cite{pro1411}. Then the OPEs will be slightly different.

Let\cite{JHEP3DBoson}
\bea
V_1(z)&=&U_1(z)=J(z)=J_1(z)+J_2(z)+\cdots+J_N(z),\label{v1zU}\\
V_2(z)&=&-h_1h_2\left( -U_2(z) + \frac{(N-1)\alpha_0}{2} U_1^\prime(z) + \frac{1}{2} (U_1 U_1)(z)\right),\label{v2zU}\\
V_3(z)&=&h_1^2h_2^2\left(U_3(z)-U_1U_2(z)+\frac{1}{3}U_1U_1U_1(z)-\frac{(N-2)\alpha_0}{2}U_2^\prime(z)\right.\nn\\
&&+\frac{\alpha_0^2(N-1)(N-2)}{12}U_1^{\prime\prime}(z)\left.+\frac{(N-1)\alpha_0}{2}U_1^\prime U_1(z)\right).\label{v3zU}
\eea
The OPEs are
\bea
V_1(z)V_{1}(w)&\sim& -\frac{N}{h_1h_2}\frac{1}{(z-w)^2},\label{v1zv1wN}\\
V_1(z)V_2(w)&\sim &\frac{V_1(w)}{(z-w)^2},\\
V_2(z)V_2(w)&\sim &\frac{c_2/2}{(z-w)^4}+\frac{2V_2{(w)}}{(z-w)^2}+\frac{V_2^\prime(w)}{z-w}\label{v2zv2wN}
\eea
with $c_2=N+h_1h_2\alpha_0^2N(N+1)(N-1),$ and
\bea
V_1(z)V_3(w)&\sim&\frac{2V_2(w)}{(z-w)^2},\\
V_2(z)V_3(w)&\sim&\frac{-h_1h_2(1+(N+1)(N-1)\alpha_0^2h_1h_2)V_1(w)}{(z-w)^4}+\frac{3V_3(w)}{(z-w)^2}+\frac{V_3^\prime(w)}{(z-w)},\label{v2zv3wN}\\
V_3(z)V_3(w)&\sim&\frac{c_3/6}{(z-w)^3}+\frac{-h_1h_2(4+(N+2)(N-2)\alpha_0^2h_1h_2)V_2(w)+\frac{3}{2}Nh_3\sigma_3U_1U_1(w)}{(z-w)^4}\nn\\
&&+\frac{-h_1h_2(4+(N+2)(N-2)\alpha_0^2h_1h_2)V_2^\prime(w)+{3}Nh_3\sigma_3U_1^\prime U_1(w)}{2(z-w)^3}+\cdots
\eea
with the central charge $c_3$
\beaa
c_3&=&-\frac{h_1h_2}{N}(N+(N+1)N(N-1)\alpha_0^2h_1h_2)(4N+(N+2)N(N-2)\alpha_0^2h_1h_2)\\
&&-3{N^2}\alpha_0^2h_1^2h_2^2.
\eeaa

The fields $V_1(z),\ V_2(z)$ and $V_3(w)$ have been already constructed in \cite{JHEP3DBoson}. In this paper, we will need the field $V_4(z)$, then we calculate it from the fact that we want the OPEs $V_1(z)V_4(w)$ and $V_2(z)V_4(w)$ are similar to that for (\ref{V4N=1}), just the coefficients will be different. We know that $V_4(z)$ is the following linear combination with the coefficients needed to be determined,
\beaa
V_4(z)&=&-h_1^3h_2^3\left(a_1 U_1'U_1U_1(z)+a_2U_1'U_1'(z)+a_3 U_1U_1U_1U_1(z)\right.\\
&&+a_4U_1''U_1(z)+a_5U_1'''(z)+a_6U_1'U_2(z)+a_7U_1U_2'(z)\\
&&+a_8U_1U_1U_2(z)+a_9U_2U_2(z)+a_{10}U_2''(z)+a_{11}U_3'(z)\\
&&\left.+a_{12}U_1U_3(z)+a_{13}U_4(z)\right),
\eeaa
From the calculations in Appendix A, we see that the coefficients $a_n,\ n=1,2\cdots,13$ satisfy the following equations
\begin{eqnarray*}
&& -\frac{24 a_{5} N}{h_{1} h_{2}} + \frac{4 a_{9} N (N-1)^{2} \alpha_0  }{h_{1}^{2} h_{2}^{2}} -\frac{12 a_{10} (N-1) N \alpha_0}{h_{1} h_{2}}
 -\frac{4 a_{11} N (N-1) (N-2) \alpha_0^{2} }{h_{1} h_{2}} \\
&& -\frac{a_{13} N (N-1) (N-2) (N-3) \alpha_0^{3} }{h_{1} h_{2}} =0,
\end{eqnarray*}

\begin{eqnarray*}
-\frac{2 a_{6} N}{h_{1} h_{2}} - \frac{2 a_{9} N (N-1) \alpha_0 }{h_{1} h_{2}} -\frac{2 a_{11} (N-2) }{h_{1} h_{2}}
-\frac{ a_{13}(N-2) (N-3) \alpha_0 }{h_{1} h_{2}}   =0,
\end{eqnarray*}

\begin{eqnarray*}
-\frac{2 a_{1} N}{h_{1} h_{2}} - \frac{2 a_{7}  (N-1)   }{h_{1} h_{2}} -\frac{a_{8} N \alpha_0 (N-1) }{h_{1} h_{2}}
-\frac{ a_{12}(N-2) (N-1) \alpha_0 }{h_{1} h_{2}}   =0,
\end{eqnarray*}

\begin{eqnarray*}
-\frac{2 a_{1} N}{h_{1} h_{2}} - \frac{ a_{6}  (N-1)   }{h_{1} h_{2}} -\frac{a_{7}(N-1) }{h_{1} h_{2}}
-\frac{  (N-1) \alpha_0 (-\frac{a_{12} N}{h_{1} h_{2}} -\frac{a_{13} (N-3)}{h_{1} h_{2}})}{2}   =0,
\end{eqnarray*}

\begin{eqnarray*}
-\frac{4 a_{3} N}{h_{1} h_{2}} - \frac{ a_{8}  (N-1)   }{h_{1} h_{2}} +\frac{a_{12}N}{3 h_{1} h_{2}}
+\frac{  a_{13} (N-3)}{3 h_{1} h_{2}}   =0,
\end{eqnarray*}

\begin{eqnarray*}
-\frac{ a_{7} N}{h_{1} h_{2}} - \frac{ a_{11}  (N-2)   }{h_{1} h_{2}} +\frac{(N-2) \alpha_0 (-\frac{a_{12} N}{h_{1} h_{2}} -\frac{a_{13} (N-3)}{h_{1} h_{2}})}{2}
 =0,
\end{eqnarray*}

\begin{eqnarray*}
-\frac{2 a_{8} N}{h_{1} h_{2}} - \frac{2 a_{9}  (N-1)   }{h_{1} h_{2}} -\frac{a_{12}(N-2)}{h_{1} h_{2}}
-\frac{a_{12} N}{h_{1} h_{2}} - \frac{ a_{13} (N-3)}{h_{1} h_{2}} =0,
\end{eqnarray*}

\begin{eqnarray*}
&& -\frac{4 a_{2} N}{h_{1} h_{2}} - \frac{6 a_{4}N   }{h_{1} h_{2}} -\frac{2 a_{6} N (N-1) \alpha_0}{h_{1} h_{2}}
 -\frac{3 a_{7} N (N-1) \alpha_0}{h_{1} h_{2}} \\
 &&+a_{9}\left(\frac{ 3 N (N-1)}{h_{1}^{2} h_{2}^{2} } + \frac{2 \alpha_0^{2} N (N-1) (N+2)}{h_{1} h_{2}}\right) +10 a_{10} (N^{3}-N )\alpha_0^{2}\\
&&+5 a_{11} \alpha_0^{3} (N+1) N (N-1) (N-2)-\frac{a_{12} N (N-1) (N-2) \alpha_0^{2} }{h_{1} h_{2}}\\
&& +\frac{3 a_{13} \alpha_0^{4} (N+1) N (N-1) (N-2) (N-3) }{2}     =0,
\end{eqnarray*}

\begin{eqnarray*}
&& -\frac{4 a_{1} N}{h_{1} h_{2}}+24 a_{5}  - \frac{2 a_{6} (N-1)   }{h_{1} h_{2}} + a_{7}\left(2 (N^{3}-N ) \alpha_0^{2} -\frac{2 (N-1)}{h_{1} h_{2}}  \right)
-\frac{2 a_{8} N \alpha_0 (N-1) }{h_{1} h_{2}}\\
&& -\frac{6 a_{9} \alpha_0 (N-1)^{2}  }{h_{1} h_{2}}  +12 \alpha_0 a_{10}  (N-1)
+2 a_{11} \alpha_0^{2} (N+3) (N-1) (N-2)\\
&& + a_{12} \left(\alpha_0^{3} (N+1) N (N-1) (N-2)- \frac{(N-1) (N-2) \alpha_0}{h_{1} h_{2}}  \right)\\
&&+ a_{13} (N-1) (N-2) (N-3) (N+2) \alpha_0^{3} =0,
\end{eqnarray*}

\begin{eqnarray*}
&& -\frac{ a_{1} N}{h_{1} h_{2}}+24 a_{5}  + \frac{a_{6} (N^{3} -N ) \alpha_0^{2}   }{2} - \frac{a_{7} (N-1)}{h_{1} h_{2}}
- \frac{4 a_{9} (N-1) N \alpha_0}{h_{1} h_{2}} +6 \alpha_0 a_{10} (N-1) \\
&& +\frac{a_{11}\alpha_0^{2} (N+3) (N-1) (N-2) }{2} +\frac{(N-1) \alpha_0}{2} \left(-\frac{ a_{8} N}{h_{1} h_{2}} + a_{9}\left((N^{3} -N) \alpha_0^{2}+\frac{8}{h_{1} h_{2}} \right)\right.\\
 &&\left. +12 a_{10} +6 a_{11} (N-2) \alpha_0-\frac{a_{12} (N-2)}{h_{1} h_{2}}
+\frac{a_{13} \alpha_0^{2} (N+5) (N-2) (N-3) }{2}   \right) =0,
\end{eqnarray*}

\begin{eqnarray*}
&& -\frac{6 a_{3} N}{h_{1} h_{2}}+6 a_{4} +3 a_{7} (N-1) \alpha_0 + a_{8}\left(\frac{(N^{3} -N) \alpha_0^{2} }{2} -\frac{2 (N-1)}{h_{1} h_{2}}\right)
-\frac{4 a_{9} (N-1) }{h_{1} h_{2}} \\
&& +\frac{a_{12} \alpha_0^{2} (N+3) (N-1) (N-2) }{2} - \frac{a_{8} N}{2 h_{1} h_{2}}
+\frac{a_{9} ((N^{3} -N) \alpha_0^{2} +\frac{8}{h_{1} h_{2}}) }{2} +6 a_{10}\\
&& +3 a_{11} (N-2) \alpha_0  -\frac{a_{12} (N-2)}{2 h_{1} h_{2}}
+\frac{a_{13} \alpha_0^{2} (N+5) (N-2) (N-3) }{4}  =0,
\end{eqnarray*}

\begin{eqnarray*}
6 a_{11} +3 a_{13} (N-3) \alpha_0 =0,
\end{eqnarray*}

\begin{eqnarray*}
2 a_{6} +4 a_{7}+2 a_{9} \alpha_0 (N-1)+2 a_{12} \alpha_0 (N-2)=0,
\end{eqnarray*}

\begin{eqnarray*}
\frac{3 a_{9} }{h_{1} h_{2}} +10 a_{10} +2 a_{11} (N-2) \alpha_0=0,
\end{eqnarray*}

\begin{eqnarray*}
2 a_{1} +a_{8} (N-1) \alpha_0 =0,
\end{eqnarray*}

\begin{eqnarray*}
12 a_{5} - \frac{ a_{9} \alpha_0 (N-1) (4 N-1) }{2 h_{1} h_{2}}+\alpha_0 a_{10} (N-1)=0,
\end{eqnarray*}

\begin{eqnarray*}
4 a_{2} +6 a_{4} + a_{6} \alpha_0 (N-1) + a_{7} (N-1) \alpha_0 -\frac{3 a_{9} (N-1)}{h_{1} h_{2}}=0.
\end{eqnarray*}
We let $a_{13}=-1$ and $a_3=\frac{1}{4}$, solve the equations above, we get
\beaa
&&a_{1} =\frac{(N-1) \alpha_0}{2},\ a_{2} =\frac{3+ \alpha_0^{2} h_1h_{2} (N-3) (2N+1 )}{20 h_{1} h_{2}},\\
&& a_{4} =\frac{2 h_{1} h_{2} N^{2} \alpha_0^{2} -5 h_{1} h_{2} N \alpha_0^{2} +7 h_{1} h_{2}  \alpha_0^{2} +5 N-7 }{20 h_{1} h_{2}},\\
&&a_{5} =\frac{ (\alpha_0^{2} h_{1} h_{2} (N-2) (N-3) +10 N-1) a (N-1)}{120 h_{1} h_{2}},\ a_{6} = -\frac{ (N-1) \alpha_0 }{2}\\
&&a_{7} = -\frac{ (N-2) \alpha_0 }{2},\ a_{8} = -1,\ a_{9} = \frac{1 }{2},\\
 &&a_{10} = -\frac{2 h_{1} h_{2} N^{2} \alpha_0^{2}-10 h_{1} h_{2} N \alpha_0^{2}+12 h_{1} h_{2} \alpha_0^{2} +3}{20 h_{1} h_{2} },\\
&&a_{11} = \frac{\alpha_0( N-3)}{2}, \ a_{12} = 1.
\eeaa
Then we have
\bea
V_4(z)&=&-h_1^3h_2^3\left(\frac{(N-1) \alpha_0}{2} U_1'U_1U_1(z)+\frac{3+ \alpha_0^{2} h_1h_{2} (N-3) (2N+1 )}{20 h_{1} h_{2}}U_1'U_1'(z)\right.\nn\\
&&+\frac{1}{4} U_1U_1U_1U_1(z)
+\frac{2 h_{1} h_{2} N^{2} \alpha_0^{2} -5 h_{1} h_{2} N \alpha_0^{2} +7 h_{1} h_{2}  \alpha_0^{2} +5 N-7 }{20 h_{1} h_{2}}U_1''U_1(z)\nn\\
&&+\frac{ (\alpha_0^{2} h_{1} h_{2} (N-2) (N-3) +10 N-1) a (N-1)}{120 h_{1} h_{2}}U_1'''(z)-\frac{ (N-1) \alpha_0 }{2}U_1'U_2(z)\nn\\
&&-\frac{ (N-2) \alpha_0 }{2}U_1U_2'(z)-U_1U_1U_2(z)+\frac{1 }{2}U_2U_2(z)\nn\\
&&-\frac{2 h_{1} h_{2} N^{2} \alpha_0^{2}-10 h_{1} h_{2} N \alpha_0^{2}+12 h_{1} h_{2} \alpha_0^{2} +3}{20 h_{1} h_{2} }U_2''(z)\nn\\
&&\left.+\frac{\alpha_0( N-3)}{2}U_3'(z)+U_1U_3(z)-U_4(z)\right).\label{v4zU}
\eea
The field $V_{n\geq 5}(z)$ can also be obtained this way. From the expressions of the fields $V_1(z),\ V_2(z),\ V_3(z)$ and $V_4(z)$, we can see some rules for general field $V_{n}(z)$, for example, it has $(-h_1h_2)^{n-1}$ as a total coefficient; In bracket, the coefficient of $U_n(z)$ is $(-1)^{n-1}$, the coefficients of terms without derivatives are constant numbers which are not dependent on $N,\ \alpha_0$ or $h_{1,2,3}$, and generally, the coefficients of terms with $k$th derivatives are polynomials of $N$ with at most $N^k$.

The field $V_4(z)$ equals
\beaa
V_4(z)&=&-h_1^3h_2^3\left(\frac{1}{4}\sum_{j=1}^NJ_jJ_jJ_jJ_j(z)-\frac{\alpha_0}{2}\sum_{j<k}J_jJ_jJ_k'(z)
-{\alpha_0}\sum_{j<k<l}(l-3)J_jJ_k'J_l(z)\right.\nn\\
&&+\frac{\alpha_0}{2}\sum_{j<k}J_jJ_k'J_k(z)+\frac{\alpha_0}{2}\sum_{j<k}J_j'J_jJ_k(z)+\frac{\alpha_0}{2}\sum_{j<k}J_j'J_kJ_k(z)\nn\\
&&-{\alpha_0}\sum_{j<k<l}(l-3)J_j'J_kJ_l(z)+\frac{\alpha_0}{2}\sum_{j=1}^N(N+1-2j)J_j'J_jJ_j(z)\\
&&-\frac{\alpha_0^2}{4}\sum_{j<k}(N+1-2k)J_jJ_k''(z)-\frac{\alpha_0^2}{4}\sum_{j<k}(N+1-2j)J_j''J_k(z)\\
&&\left.-\alpha_0^2\sum_{j<k}(k-j)J_j'J_k'(z)+\frac{3}{20h_1h_2}\sum_{j=1}^NJ_j'J_j'(z)-\frac{1}{10h_1h_2}\sum_{j=1}^NJ_j''J_j(z)+\cdots\right).
\eeaa
When $N=1, h_1=1,h_2=-1$, $V_4(z)$ becomes \[\frac{1}{4}:J_1(z)^4:-\frac{3}{20}:J_1'(z)^2:+\frac{1}{10}:J_1''(z)J_1(z):,\]
which is the same with $V_4(z)$ in (\ref{V4N=1}). We have
\bea
V_1(z)V_4(w)&\sim&{\frac {3h_1h_2(-1+{\alpha_0}^{2}h_1h_{2}\left( {N}^{2}+1 \right))}{5}}\frac{V_1(w)}{(z-w)^4}\nn\\
&&-{\frac {3h_1h_2(-1+{\alpha_0}^{2}h_1h_{2}\left( {N}^{2}+1 \right))}{5}}\frac{V_1'(w)}{(z-w)^3}\nn\\
&&+\frac{3V_3(w)}{(z-w)^2}+\frac{h_1h_2(-1+{\alpha_0}^{2}h_1h_{2}\left( {N}^{2}+1 \right))}{10}\frac{V_1''(w)}{(z-w)^2},\label{v1zv4wN}\\
V_2(z)V_4(w)&\sim&\frac{-h_1h_2(21+\alpha_0^2h_1h_2(9N^2-21))V_2(w)}{5(z-w)^4}+\frac{4V_4(w)}{(z-w)^2}+\frac{V_4'(w)}{(z-w)}.\label{v2zv4wN}
\eea
\section{Schur functions, KP hierarchy and $W_{1+\infty}$ algebra}\label{sect3}
In this section, we realize the KP hierarchy from the special case of the $W_{1+\infty}$ algebra introduced in the last section. We see that when $\alpha_0=-h_3/h_1h_2$ and $N=1, h_1=1,h_2=-1$, the fields $V_j(z)$ constructed from the Miura transformation become that represented by Boson field $B(z)$, and the mode $V_{j,m}$ satisfy the relations in (\ref{vjmvkn}).

We know that $V_j(z)$ in (\ref{V4N=1}) satisfy
\beaa
V_2(z)V_1(w)&\sim&\frac{V_1(w)}{(z-w)^2}+\frac{V_1'(w)}{z-w},\\
{V}_2(z){V}_2(w)&\sim& \frac{1/2}{(z-w)^4}+\frac{2{V}_2(w)}{(z-w)^2}+\frac{{V}_2^\prime(w)}{(z-w)},\\
{V}_2(z){V}_3(w)&\sim &\frac{{V}_1(w)}{(z-w)^4}+\frac{3{V}_3(w)}{(z-w)^2}+\frac{{V}_3^\prime(w)}{(z-w)},\\
{V}_2(z){V}_4(w)&\sim &\frac{21}{5}\frac{{V}_2(w)}{(z-w)^4}+\frac{4{V}_4(w)}{(z-w)^2}+\frac{{V}_4^\prime(w)}{(z-w)}.
\eeaa
Let's treat $V_2(z)$ as the generator of the conformal transformations. Consider
\be
z=e^{ix}
\ee
with $i=\sqrt{-1}$, under the conformal transformations, the field $V_j(z)$ transform to $\tilde{V}_j(x)$ as
\bea
\tilde{V}_1(x)&=&\left(\frac{\text{d}x}{\text{d}z}\right)^{-1}V_1(z)=i\sum_{m\in\Z}V_{1,m}e^{-imx},\\
\tilde{V}_2(x)&=&-\sum_{m\in\Z}V_{2,m}e^{-imx}+\frac{1}{24},\\
\tilde{V}_3(x)&=&-i\sum_{m\in\Z}V_{3,m}e^{-imx}+\frac{i}{12}\sum_{m\in\Z}V_{1,m}e^{-imx},\\
\tilde{V}_4(x)&=&\sum_{m\in\Z}V_{4,m}e^{-imx}-\frac{7}{20}\sum_{m\in\Z}V_{2,m}e^{-imx}.
\eea
Others can be calculated this way. Define
\be
v_n=(-i)^{n+1}\tilde{V}_{n+1}(x), \ n=0,1,2,\cdots.
\ee
Let $\{,\}=-i[,]$, the Poisson bracket relations can be calculated,
\beaa
&&\{v_0(x),v_3(y)\}=-i[v_0(x),v_3(y)]\\
&=&-i[\sum_{m\in\Z}V_{1,m}e^{-imx},\sum_{n\in\Z}V_{4,n}e^{-iny}-\frac{7}{20}\sum_{n\in\Z}V_{2,n}e^{-iny}]\\
&=&-i\sum_{m,n}\left(\frac{1}{2}(m+n)^3-(m+n)^2n+\frac{3}{5}(m+n)n^2-\frac{1}{10}n^3-\frac{1}{4}(m+n)+\frac{1}{4}n\right)\\
&&\cdot V_{1,m+n}e^{-i(m+n)x}e^{in(x-y)}\\
&&-i\sum_{m,n}3(m+n)V_{3,m+n}e^{-i(m+n)x}e^{in(x-y)}-i\sum_{m,n}(-3n)V_{3,m+n}e^{-i(m+n)x}e^{in(x-y)}\\
&=&-\frac{1}{2}v_{0,xxx}\delta(x-y)-v_{0,xx}\partial_x\delta(x-y)-\frac{3}{5}v_{0,x}\partial_x^2\delta(x-y)
-\frac{1}{10}v_{0}\partial_x^3\delta(x-y)\\
&&+3v_{2,x}\delta(x-y)+3v_{2}\partial_x\delta(x-y).
\eeaa
In this way, we also obtain
\beaa
\{v_0(x),v_0(y)\}&=&\partial_x\delta(x-y),\\
\{v_0(x),v_1(y)\}&=&v_{0,x}\delta(x-y)+v_{0}\partial_x\delta(x-y),\\
\{v_0(x),v_2(y)\}&=&2v_{1,x}\delta(x-y)+2v_{1}\partial_x\delta(x-y),\\
\{v_1(x),v_2(y)\}&=&2v_{2,x}\delta(x-y)+3v_{2}\partial_x\delta(x-y)-\frac{1}{6}v_{0,xxx}\delta(x-y)\\
&&-\frac{1}{2}v_{0,xx}\partial_x\delta(x-y)-\frac{1}{2}v_{0,x}\partial_x^2\delta(x-y)-\frac{1}{6}v_{0}\partial_x^3\delta(x-y).
\eeaa

Define the pseudo-differential operator $L$ by
\be
L=\partial+\sum_{n=0}^\infty v_n \partial^{-n-1}.
\ee
The Hamiltonians are given by
\be
H_n=\frac{1}{n+1}\int\text{Res}L^{n+1}\text{d}x,
\ee
we list the first few terms:
\beaa
H_0&=&\int v_0\text{d}x,\\
H_1&=&\int v_1\text{d}x,\\
H_2&=&\int (v_2+v_0^2)\text{d}x,\\
H_3&=&\int (v_3+3v_0v_1)\text{d}x.\\
\eeaa
Note that the expressions of $L$ and $H_n$ here are the same with that in \cite{Minru}, but the fields $v_n$ are different.

From the Poisson evolution equation
\be
\frac{\partial v_i(t,x)}{\partial t_n}=\{v_i,H_n\},
\ee
the KP hierarchy is obtained. The first member of the KP hierarchy is called the KP equation. We calculate the KP equation in the following:
\bea
\frac{\partial v_0}{\partial t_3}&=&-\frac{1}{2}v_{0,xxx}+3v_{2,x}+6v_0v_{0,x}+3v_{1,x},\label{v0t3}\\
\frac{\partial v_1}{\partial t_2}&=&2v_{2,x}-\frac{1}{6}v_{0,xxx}+2v_0v_{0,x},\label{v1t2}\\
\frac{\partial v_0}{\partial t_2}&=&2v_{1,x}+2v_{0,x},\label{v0t2},
\eea
where the subscript $x$ denotes the derivative with respect to the variable $x$. From (\ref{v1t2}) and (\ref{v0t2}), we obtain
\beaa
v_1&=&\frac{1}{2}\partial_x^{-1}v_{0,t_2}-v_0,\\
v_2&=&\frac{1}{4}\partial_x^{-2}v_{0,t_2t_2}-\frac{1}{2}\partial_x^{-1}v_{0,t_2}+\frac{1}{12}v_{0,xx}-\frac{1}{2}v_0^2,\\
\eeaa
substituting them into (\ref{v0t3}), we obtain
\be
v_{0,t_3}=-\frac{1}{4}v_{0,xxx}+\frac{3}{4}\partial_x^{-1}v_{0,t_2t_2}+3v_0v_{0,x}-3v_{0,x}.
\ee
Let
\be
\frac{\partial}{\partial t}:=\frac{\partial}{\partial t_3}+3\frac{\partial}{\partial x},
\ee
which corresponds to replacing the Hamiltonian $H_3$ by $H_3+3H_1$, then the equation above becomes
\[
v_{0,t}=-\frac{1}{4}v_{0,xxx}+\frac{3}{4}\partial_x^{-1}v_{0,t_2t_2}+3v_0v_{0,x},
\]
Let $y=t_2$, then we have,
\be
\frac{3}{4}v_{0,yy}=(v_{0,t}+\frac{1}{4}v_{0,xxx}-3v_0v_{0,x})_x,
\ee
this is the KP equation. Multiplied $v_0,x,y,t$ by a constant respectively and still denoted by $v_0,x,y,t$, the KP equation becomes
\be
\frac{3}{4}v_{0,yy}=(v_{0,t}-\frac{1}{4}v_{0,xxx}-\frac{3}{2}v_0v_{0,x})_x.
\ee
This equation is still called the KP equation. Let $u=2(\text{log}\tau)_{xx}$, we get
\[
3\tau_{yy}\tau-3\tau_y\tau_y-4\tau_{tx}\tau+4\tau_t\tau_x+\tau_{xxxx}\tau-4\tau_{xxx}\tau_x+3\tau_{xx}\tau_{xx}=0,
\]
which can also be expressed as
\be
S_{\begin{tikzpicture}
\draw [step=0.2](0,0) grid(.4,.2);
\draw [step=0.2](0,-0.2) grid(.4,0);
\end{tikzpicture}}^\bot\tau\cdot\tau-S_{\begin{tikzpicture}
\draw [step=0.2](0,0) grid(.4,.2);
\draw [step=0.2](0,-0.2) grid(.2,0);
\end{tikzpicture}}^\bot\tau\cdot S_{\begin{tikzpicture}
\draw [step=0.2](0,0) grid(.2,.2);
\end{tikzpicture}}^\bot\tau+S_{\begin{tikzpicture}
\draw [step=0.2](0,0) grid(.4,.2);
\end{tikzpicture}}^\bot\tau\cdot S_{\begin{tikzpicture}
\draw [step=0.2](0,0) grid(.2,.4);
\end{tikzpicture}}^\bot\tau=0,\label{kpequationS}
\ee
where $S_{\lambda}$ is the Schur functions, and $S_{\lambda}^\bot$ is obtained from Schur function $S_\lambda$ by replacing $x_n$ by $\frac{1}{n}\frac{\partial}{\partial x_n}$ with $x_1=x,x_2=y,x_3=t$.

Write the tau functions of the KP hierarchy in the form
\[
\tau=\sum_{\lambda}c_{\lambda}S_{\lambda},
\]
we know that the coefficients $c_\lambda$ satisfy the Pl\"ucker relations. From (\ref{kpequationS}), we see that the first Pl\"ucker relation is
\[
c_{\begin{tikzpicture}
\draw [step=0.2](0,0) grid(.4,.2);
\draw [step=0.2](0,-0.2) grid(.4,0);
\end{tikzpicture}}c_{\emptyset}-c_{\begin{tikzpicture}
\draw [step=0.2](0,0) grid(.4,.2);
\draw [step=0.2](0,-0.2) grid(.2,0);
\end{tikzpicture}}c_{\begin{tikzpicture}
\draw [step=0.2](0,0) grid(.2,.2);
\end{tikzpicture}}+c_{\begin{tikzpicture}
\draw [step=0.2](0,0) grid(.4,.2);
\end{tikzpicture}}c_{\begin{tikzpicture}
\draw [step=0.2](0,0) grid(.2,.4);
\end{tikzpicture}}=0.
\]
\section{Jack polynomials, KP hierarchy and $W_{1+\infty}$ algebra}\label{sect4}
In this section, we construct the deformed KP hierarchy from the $W_{1+\infty}$ algebra introduced in section 3. Parallel to that the tau functions of the KP hierarchy can be represented by the linear combinations of Schur functions, we want that the tau functions of the KP hierarchy can be represented by the linear combinations of Jack polynomials.

Let $\alpha_0=-h_3/h_1h_2$ and $N=1$, the fields
\bea
{V}_1(z)&=&J_1(z),\label{v1N=1J}\nn\\
{V}_2(z)&=&-h_1h_2\frac{1}{2}:J_1(z)^2:,\nn\\
{V}_3(z)&=&h_1^2h_2^2\frac{1}{3}:J_1(z)^3:,\label{vN=1J}\\
{V}_4(z)&=&-h_1^3h_2^3\left(\frac{1}{4}:J_1(z)^4:+\frac{3-6\alpha_0^2h_1h_2}{20h_1h_2}:J_1'(z)^2:+\frac{2\alpha_0^2h_1h_2-1}{10}:J_1''(z)J_1(z):\right).\nn
\eea
They satisfy the following OPEs
\beaa
V_1(z)V_2(w)&\sim&\frac{V_1(w)}{(z-w)^2},\\
{V}_1(z){V}_3(w)&\sim& \frac{2V_2(w)}{(z-w)^2},\\
{V}_1(z){V}_4(w)&\sim &\left(-\frac{3h_1h_2}{5}+\frac{6\alpha_0^2h_1^2h_2^2}{5}\right)\frac{{V}_1(w)}{(z-w)^4}+\left(\frac{3h_1h_2}{5}
-\frac{6\alpha_0^2h_1^2h_2^2}{5}\right)\frac{{V}_1'(w)}{(z-w)^3}\\
&&+\left(3V_3(w)+\left(-\frac{h_1h_2}{10}+\frac{\alpha_0^2h_1^2h_2^2}{5}\right)V_1''(w)\right)\frac{1}{(z-w)^2},
\eeaa
and
\beaa
V_2(z)V_1(w)&\sim&\frac{V_1(w)}{(z-w)^2}+\frac{V_1'(w)}{(z-w)},\\
{V}_2(z){V}_2(w)&\sim& \frac{1/2}{(z-w)^4}+\frac{2V_2(w)}{(z-w)^2}+\frac{V_2'(w)}{(z-w)},\\
{V}_2(z){V}_3(w)&\sim &\frac{-h_1h_2V_1(w)}{(z-w)^4}+\frac{3V_3(w)}{(z-w)^3}+\frac{V_3'(w)}{(z-w)^3},\\ {V}_2(z){V}_4(w)&\sim &\left(-\frac{21h_1h_2}{5}+\frac{12\alpha_0^2h_1^2h_2^2}{5}\right)\frac{{V}_2(w)}{(z-w)^4}+\frac{4V_4(w)}{(z-w)^2}+\frac{V_4'(w)}{(z-w)},
\eeaa
and
\beaa
&&V_3(z)V_3(w)\sim-\frac{2}{3}h_1h_2\frac{1}{(z-w)^6}+\frac{-h_1h_24V_2(w)}{(z-w)^4}+\frac{-h_1h_22V_2'(w)}{(z-w)^3}\\
&&+\left(4V_4(w)-h_1h_2\frac{3}{5}V_2''(w)+\frac{2}{5}\alpha_0^2h_1^3h_2^3(-3J_1'J_1'(w)+2J_1''J_1(w))\right)\frac{1}{(z-w)^2}\\
&&+\left(2V_4'(w)-h_1h_2\frac{2}{15}V_2'''(w)+\frac{2}{5}\alpha_0^2h_1^3h_2^3(-2J_1''J_1'(w)+J_1'''J_1(w))\right)\frac{1}{(z-w)^2}.
\eeaa
We can see that when $h_1=1, h_2=-1$, we have $\alpha_0=0$ and the fields $V_j(z)$ and the OPE relations between them become that for (\ref{V4N=1}). The commutation relations are
\beaa
[V_{1,n},V_{1,m}]=-\frac{N}{h_1h_2}n\delta_{n+m,0},\
[V_{1,n},V_{2,m}]= n V_{1,m+n},\
[V_{1,n},V_{3,m}]=2nV_{2,m+n},\eeaa
\beaa
[V_{1,m},V_{4,n}]&=&\left(-\frac{h_1h_2}{10}+\frac{\alpha_0^2h_1^2h_2^2}{5}\right)\left(5m^3+5m^2n+mn^2+m\right)V_{1,m+n}+3mV_{3,m+n},\\
\eeaa
and
\beaa
[V_{2,m},V_{3,n}]=-h_1h_2\frac{1}{6}(m^3-m)V_{1,m+n}+(2m-n)V_{3,m+n}.
\eeaa

We treat $V_2(z)$ as the generator of the conformal transformations. Consider
\be
z=e^{ix},
\ee
under the conformal transformations, the field $V_j(z)$ transform to $\tilde{V}_j(x)$ as
\bea
\tilde{V}_1(x)&=&i\sum_{m\in\Z}V_{1,m}e^{-imx},\\
\tilde{V}_2(x)&=&-\sum_{m\in\Z}V_{2,m}e^{-imx}+\frac{1}{24},\\
\tilde{V}_3(x)&=&-i\sum_{m\in\Z}V_{3,m}e^{-imx}-\frac{ih_1h_2}{12}\sum_{m\in\Z}V_{1,m}e^{-imx},\\
\tilde{V}_4(x)&=&z^4V_4(z)-\left(-\frac{7h_1h_2}{20}+\frac{\alpha_0^2h_1^2h_2^2}{5}\right)z^2V_2(z)\nn\\
&=&\sum_{m\in\Z}V_{4,m}e^{-imx}-\left(-\frac{7h_1h_2}{20}+\frac{\alpha_0^2h_1^2h_2^2}{5}\right)\sum_{m\in\Z}V_{2,m}e^{-imx}.
\eea
Define
\be
v_n=(-i)^{n+1}\tilde{V}_{n+1}(x), \ n=0,1,2,\cdots.
\ee
Let $\{,\}=-i[,]$, the Poisson bracket relations can be calculated,
\beaa
\{v_0(x),v_3(y)\}
&=&\left(-\frac{h_1h_2}{10}+\frac{\alpha_0^2h_1^2h_2^2}{5}\right)\left(-5v_{0,xxx}\delta(x-y)
-10v_{0,xx}\partial_x\delta(x-y)\right.\\
&&\left.-6v_{0,x}\partial_x^2\delta(x-y)
-v_{0}\partial_x^3\delta(x-y)\right)\\
&&+3v_{2,x}\delta(x-y)+3v_{2}\partial_x\delta(x-y),\\
\{v_0(x),v_0(y)\}&=&-\frac{1}{h_1h_2}\partial_x\delta(x-y),\\
\{v_0(x),v_1(y)\}&=&v_{0,x}\delta(x-y)+v_{0}\partial_x\delta(x-y),\\
\{v_0(x),v_2(y)\}&=&2v_{1,x}\delta(x-y)+2v_{1}\partial_x\delta(x-y),\\
\{v_1(x),v_2(y)\}&=&2v_{2,x}\delta(x-y)+3v_{2}\partial_x\delta(x-y)+\frac{h_1h_2}{6}v_{0,xxx}\delta(x-y)\\
&&+\frac{h_1h_2}{2}v_{0,xx}\partial_x\delta(x-y)+\frac{h_1h_2}{2}v_{0,x}\partial_x^2\delta(x-y)+\frac{h_1h_2}{6}v_{0}\partial_x^3\delta(x-y).
\eeaa
Define the pseudo-differential operator $L$ by
\be
L=\partial+\sum_{n=0}^\infty v_n \partial^{-n-1},
\ee
and the Hamiltonians by
\be
H_n=\frac{1}{n+1}\int\text{Res}L^{n+1}\text{d}x,
\ee
as in the last section. The KP hierarchy is realized as the Poisson evolution equation
\be
\frac{\partial v_i(t,x)}{\partial t_n}=\{v_i,H_n\},
\ee
and the KP equation is obtained from
\bea
\frac{\partial v_0}{\partial t_3}&=&\left(\frac{h_1h_2}{2}-\alpha_0^2h_1^2h_2^2\right)v_{0,xxx}+3v_{2,x}+6v_0v_{0,x}-\frac{1}{h_1h_2}3v_{1,x},\label{v0t3J}\\
\frac{\partial v_1}{\partial t_2}&=&2v_{2,x}+\frac{h_1h_2}{6}v_{0,xxx}+2v_0v_{0,x},\label{v1t2J}\\
\frac{\partial v_0}{\partial t_2}&=&2v_{1,x}-\frac{1}{h_1h_2}2v_{0,x},\label{v0t2J}
\eea
From (\ref{v1t2J}) and (\ref{v0t2J}), we obtain
\beaa
v_1&=&\frac{1}{2}\partial_x^{-1}v_{0,t_2}+\frac{1}{h_1h_2}v_0,\\
v_2&=&\frac{1}{4}\partial_x^{-2}v_{0,t_2t_2}+\frac{1}{2h_1h_2}\partial_x^{-1}v_{0,t_2}-\frac{h_1h_2}{12}v_{0,xx}-\frac{1}{2}v_0^2,\\
\eeaa
substituting them into (\ref{v0t3J}), we obtain
\be
v_{0,t_3}=\left(\frac{h_1h_2}{4}-\alpha_0^2h_1^2h_2^2\right)v_{0,xxx}+\frac{3}{4}\partial_x^{-1}v_{0,t_2t_2}+3v_0v_{0,x}-\frac{1}{h_1^2h_2^2}3v_{0,x}.
\ee
Let
\be
\frac{\partial}{\partial t}:=\frac{\partial}{\partial t_3}+\frac{1}{h_1^2h_2^2}3\frac{\partial}{\partial x},
\ee
which corresponds to replacing the Hamiltonian $H_3$ by $H_3+\frac{1}{h_1^2h_2^2}3H_1$, then the equation above becomes
\[
v_{0,t}=\left(\frac{h_1h_2}{4}-\alpha_0^2h_1^2h_2^2\right)v_{0,xxx}+\frac{3}{4}\partial_x^{-1}v_{0,t_2t_2}+3v_0v_{0,x},
\]
Let $y=t_2$, then we have,
\be
\frac{3}{4}v_{0,yy}=\left(v_{0,t}-\left(\frac{h_1h_2}{4}-\alpha_0^2h_1^2h_2^2\right)v_{0,xxx}-3v_0v_{0,x}\right)_x,
\ee
this is the KP equation. Clearly, it becomes the KP hierarchy in the last section when $h_1=1,h_2=-1$. Multiplied $v_0,x,y,t$ by a constant respectively and still denoted by $v_0,x,y,t$, the KP equation becomes
\be
\frac{3}{4}v_{0,yy}=\left(v_{0,t}+\left(\frac{h_1h_2}{4}-\alpha_0^2h_1^2h_2^2\right)v_{0,xxx}-\frac{3}{2}v_0v_{0,x}\right)_x.
\ee
Let
$$u=-(h_1h_2-4h_3^2)2(\text{log}\tau)_{xx},$$
 we get
\beaa
&&3\tau_{yy}\tau-3\tau_y\tau_y-4\tau_{tx}\tau+4\tau_t\tau_x\\
&&-\left(h_1h_2-4h_3^2\right)\left(\tau_{xxxx}\tau-4\tau_{xxx}\tau_x+3\tau_{xx}\tau_{xx}\right)=0.
\eeaa
The Jack polynomials $Y_\lambda$ are recalled in Appendix B, we use $Y_\lambda^\bot$ to denote the operators obtained from $Y_\lambda$ by replacing $x_n$ by $\frac{1}{n}\frac{\partial}{\partial x_n}$, similar to that in the case for Schur functions. Then the deformed KP equation can be written as
\bea
&&(3h_1+4h_2)(h_1+h_2)\left(Y_{\begin{tikzpicture}
\draw [step=0.2](0,0) grid(.8,.2);
\end{tikzpicture}}^\bot\tau\right)\cdot\tau+(3h_2+4h_1)(h_1+h_2)\left(Y_{\begin{tikzpicture}
\draw [step=0.2](0,0) grid(.2,.8);
\end{tikzpicture}}^\bot\tau\right)\cdot\tau\nn\\
&&+\frac{2(3h_1+4h_2)(h_1+h_2)(2h_1-h_2)}{h_1-h_2}\left(Y_{\begin{tikzpicture}
\draw [step=0.2](0,0) grid(.6,.2);
\draw [step=0.2](0,-0.2) grid(.2,0);
\end{tikzpicture}_{h_1,2h_1,h_2}}^\bot\tau\right)\cdot\tau\nn\\
&&+\frac{2(3h_2+4h_1)(h_1+h_2)(2h_2-h_1)}{h_2-h_1}\left(Y_{\begin{tikzpicture}
\draw [step=0.2](0,0) grid(.4,.2);
\draw [step=0.2](0,-0.4) grid(.2,0);
\end{tikzpicture}_{h_2,2h_2,h_1}}^\bot\tau\right)\cdot\tau\nn\\
&&+\frac{3(5h_1^2+4h_1h_2+5h_2^2)(h_1-h_2)}{h_1-2h_2}\left(Y_{\begin{tikzpicture}
\draw [step=0.2](0,0) grid(.4,.2);
\draw [step=0.2](0,-0.2) grid(.4,0);
\end{tikzpicture}_{h_1,h_2,h_1+h_2}}^\bot\tau\right)\cdot\tau\nn\\
&&
-4(3h_1+4h_2)(h_1+h_2)\left(Y_{\begin{tikzpicture}
\draw [step=0.2](0,0) grid(.6,.2);
\end{tikzpicture}}^\bot\tau\right)\cdot \left(Y_{\begin{tikzpicture}
\draw [step=0.2](0,0) grid(.2,.2);
\end{tikzpicture}}^\bot\tau\right)\nn\\
&&-\frac{6(8h_1^2+13h_1h_2+8h_2^2)(h_1-h_2)}{h_1-2h_2}\left(Y_{\begin{tikzpicture}
\draw [step=0.2](0,0) grid(.4,.2);
\draw [step=0.2](0,-0.2) grid(.2,0);
\end{tikzpicture}_{h_1,h_2}}^\bot\tau\right)\cdot \left(Y_{\begin{tikzpicture}
\draw [step=0.2](0,0) grid(.2,.2);
\end{tikzpicture}}^\bot\tau\right)\nn\\
&&-4(3h_2+4h_1)(h_1+h_2)\left(Y_{\begin{tikzpicture}
\draw [step=0.2](0,0) grid(.2,.6);
\end{tikzpicture}}^\bot\tau\right)\cdot \left(Y_{\begin{tikzpicture}
\draw [step=0.2](0,0) grid(.2,.2);
\end{tikzpicture}}^\bot\tau\right)\nn\\
&&+3(3h_1+4h_2)(h_1+h_2)\left(Y_{\begin{tikzpicture}
\draw [step=0.2](0,0) grid(.4,.2);
\end{tikzpicture}}^\bot\tau\right)\cdot \left(Y_{\begin{tikzpicture}
\draw [step=0.2](0,0) grid(.4,.2);
\end{tikzpicture}}^\bot\tau\right)\nn\\
&&+12(h_1^2+h_1h_2+h_2^2)\left(Y_{\begin{tikzpicture}
\draw [step=0.2](0,0) grid(.4,.2);
\end{tikzpicture}}^\bot\tau\right)\cdot \left(Y_{\begin{tikzpicture}
\draw [step=0.2](0,0) grid(.2,.4);
\end{tikzpicture}}^\bot\tau\right)\nn\\
&&+3(3h_2+4h_1)(h_1+h_2)\left(Y_{\begin{tikzpicture}
\draw [step=0.2](0,0) grid(.2,.4);
\end{tikzpicture}}^\bot\tau\right)\cdot \left(Y_{\begin{tikzpicture}
\draw [step=0.2](0,0) grid(.2,.4);
\end{tikzpicture}}^\bot\tau\right)=0.
\label{kpequationJ}
\eea
Write the tau functions of the deformed KP hierarchy in the form
\[
\tau=\sum_{\lambda}c_{\lambda}Y_{\lambda},
\]
we see that the coefficients $c_\lambda$ satisfy the Pl\"ucker relations. From (\ref{kpequationS}), we see that the first Pl\"ucker relation is
\bea
&&(3h_1+4h_2)(h_1+h_2)c_{\begin{tikzpicture}
\draw [step=0.2](0,0) grid(.8,.2);
\end{tikzpicture}}c_{\emptyset}+(3h_2+4h_1)(h_1+h_2)c_{\begin{tikzpicture}
\draw [step=0.2](0,0) grid(.2,.8);
\end{tikzpicture}}c_{\emptyset}\nn\\
&&+\frac{2(3h_1+4h_2)(h_1+h_2)(2h_1-h_2)}{h_1-h_2}c_{\begin{tikzpicture}
\draw [step=0.2](0,0) grid(.6,.2);
\draw [step=0.2](0,-0.2) grid(.2,0);
\end{tikzpicture}_{h_1,2h_1,h_2}}c_{\emptyset}\nn\\
&&+\frac{2(3h_2+4h_1)(h_1+h_2)(2h_2-h_1)}{h_2-h_1}c_{\begin{tikzpicture}
\draw [step=0.2](0,0) grid(.4,.2);
\draw [step=0.2](0,-0.4) grid(.2,0);
\end{tikzpicture}_{h_2,2h_2,h_1}}c_\emptyset\nn\\
&&+\frac{3(5h_1^2+4h_1h_2+5h_2^2)(h_1-h_2)}{h_1-2h_2}c_{\begin{tikzpicture}
\draw [step=0.2](0,0) grid(.4,.2);
\draw [step=0.2](0,-0.2) grid(.4,0);
\end{tikzpicture}_{h_1,h_2,h_1+h_2}}c_\emptyset\nn\\
&&
-4(3h_1+4h_2)(h_1+h_2)c_{\begin{tikzpicture}
\draw [step=0.2](0,0) grid(.6,.2);
\end{tikzpicture}}c_{\begin{tikzpicture}
\draw [step=0.2](0,0) grid(.2,.2);
\end{tikzpicture}}\nn\\
&&-\frac{6(8h_1^2+13h_1h_2+8h_2^2)(h_1-h_2)}{h_1-2h_2}c_{\begin{tikzpicture}
\draw [step=0.2](0,0) grid(.4,.2);
\draw [step=0.2](0,-0.2) grid(.2,0);
\end{tikzpicture}_{h_1,h_2}}c_{\begin{tikzpicture}
\draw [step=0.2](0,0) grid(.2,.2);
\end{tikzpicture}}\nn\\
&&-4(3h_2+4h_1)(h_1+h_2)c_{\begin{tikzpicture}
\draw [step=0.2](0,0) grid(.2,.6);
\end{tikzpicture}}c_{\begin{tikzpicture}
\draw [step=0.2](0,0) grid(.2,.2);
\end{tikzpicture}}+3(3h_1+4h_2)(h_1+h_2)c_{\begin{tikzpicture}
\draw [step=0.2](0,0) grid(.4,.2);
\end{tikzpicture}}c_{\begin{tikzpicture}
\draw [step=0.2](0,0) grid(.4,.2);
\end{tikzpicture}}\nn\\
&&+12(h_1^2+h_1h_2+h_2^2)c_{\begin{tikzpicture}
\draw [step=0.2](0,0) grid(.4,.2);
\end{tikzpicture}}c_{\begin{tikzpicture}
\draw [step=0.2](0,0) grid(.2,.4);
\end{tikzpicture}}+3(3h_2+4h_1)(h_1+h_2)c_{\begin{tikzpicture}
\draw [step=0.2](0,0) grid(.2,.4);
\end{tikzpicture}}c_{\begin{tikzpicture}
\draw [step=0.2](0,0) grid(.2,.4);
\end{tikzpicture}}=0.\label{kpequationJc}
\eea
Clearly, when $h_1=1,h_2=-1,$ the KP equation (\ref{kpequationJ}) becomes the KP equation (\ref{kpequationS}), the Pl\"ucker relation (\ref{kpequationJc}) becomes the first Pl\"ucker relation for KP hierarchy.

Note that the results in this section are all symmetric about $x$-axis and $y$-axis, that is, they are symmetric about $h_1$ and $h_2$. If we take $h_1=\sqrt{\alpha}, h_2=-1/\sqrt{\alpha}$, the results become that for the usual Jack polynomials.
\section{3-Jack polynomials, KP hierarchy and $W_{1+\infty}$ algebra }\label{sect5}
We consider 3D Young diagrams which have at most $N$ layers in the $z$-axis direction. The field $J_j(z)$ is the Bosonic field corresponding to 2D Young diagrams on the slice $z=j$ of 3D Young diagrams for $j=1,2,\cdots,N$. The fields\cite{JHEP3DBoson}
\bea
V_1(z)&=&J_1(z)+J_2(z)+\cdots+J_N(z),\\
V_2(z)&=&-\frac{h_1h_2}{2}\sum_{j=1}^N :J_j(z)J_j(z):+\frac{h_3}{2}\sum_{j=1}^N(N+1-2j) J_j^\prime(z),\label{v2zj}\\
V_3(z)&=&-\frac{1}{3}h_1^2h_2^2\sum_{j=1}^N:J_1(z)^3:+\frac{1}{2}\alpha_0h_1^2h_2^2\sum_{j<k}J_jJ_k^\prime(z)-\frac{1}{2}\alpha_0h_1^2h_2^2\sum_{j<k}J_j^\prime J_k(z)\nn\\
&&-\frac{1}{2}\alpha_0h_1^2h_2^2\sum_{j=1}^N(N+1-2j)J_j^\prime J_j(z)\nn\\
&&+\alpha_0^2h_1^2h_2^2\sum_{j=1}^N\left(\frac{(j-1)(N-j)}{2}-\frac{(N-1)(N-j)}{12}\right)J_j^{\prime\prime}(z),
\eea
they equal $V_j(z)$ in (\ref{v1zU}-\ref{v3zU}) respectively, and $V_4(z)$ is introduced in (\ref{v4zU}). The OPEs between them have been given in section 3. The commutation relations are
\beaa
[V_{1,n},V_{1,m}]=-\frac{N}{h_1h_2}n\delta_{n+m,0},\
[V_{1,n},V_{2,m}]= n V_{1,m+n},\
[V_{1,n},V_{3,m}]=2nV_{2,m+n},\eeaa
\beaa
[V_{1,m},V_{4,n}]&=&\frac {h_1h_2(-1+{\alpha_0}^{2}h_1h_{2}\left( {N}^{2}+1 \right))}{10}
\left(5m^3+5m^2n+mn^2+m\right)V_{1,m+n}\\
&&+3mV_{3,m+n},\\
\eeaa
and
\beaa
[V_{2,m},V_{3,n}]=-h_1h_2(1+(N+1)(N-1)\alpha_0^2h_1h_2)\frac{1}{6}(m^3-m)V_{1,m+n}+(2m-n)V_{3,m+n}.
\eeaa

Similarly, we treat $V_2(z)$ as the generator of the conformal transformations. Consider
\be
z=e^{ix},
\ee
under the conformal transformations, the field $V_j(z)$ transform to $\tilde{V}_j(x)$ as
\bea
\tilde{V}_1(x)&=&i\sum_{m\in\Z}V_{1,m}e^{-imx},\label{tildev13J}\nn\\
\tilde{V}_2(x)&=&-\sum_{m\in\Z}V_{2,m}e^{-imx}+\frac{N+h_1h_2\alpha_0^2N(N+1)(N-1)}{24},\nn\\
\tilde{V}_3(x)&=&-i\sum_{m\in\Z}V_{3,m}e^{-imx}-\frac{ih_1h_2(1+(N+1)(N-1)\alpha_0^2h_1h_2)}{12}\sum_{m\in\Z}V_{1,m}e^{-imx},\nn\\
\tilde{V}_4(x)&=&z^4V_4(z)+\frac{h_1h_2(7+\alpha_0^2h_1h_2(3N^2-7))}{20}z^2V_2(z)\nn\\
&=&\sum_{m\in\Z}V_{4,m}e^{-imx}+\frac{h_1h_2(7+\alpha_0^2h_1h_2(3N^2-7))}{20}\sum_{m\in\Z}V_{2,m}e^{-imx}.
\eea
Define
\be
v_n=(-i)^{n+1}\tilde{V}_{n+1}(x), \ n=0,1,2,\cdots.
\ee
Let $\{,\}=-i[,]$, the Poisson bracket relations can be calculated,
\beaa
\{v_0(x),v_3(y)\}
&=&\frac {h_1h_2(-1+{\alpha_0}^{2}h_1h_{2}\left( {N}^{2}+1 \right))}{10}\left(-5v_{0,xxx}\delta(x-y)
-10v_{0,xx}\partial_x\delta(x-y)\right.\\
&&\left.-6v_{0,x}\partial_x^2\delta(x-y)
-v_{0}\partial_x^3\delta(x-y)\right)+3v_{2,x}\delta(x-y)+3v_{2}\partial_x\delta(x-y),\\
\{v_0(x),v_0(y)\}&=&-\frac{N}{h_1h_2}\partial_x\delta(x-y),\\
\{v_0(x),v_1(y)\}&=&v_{0,x}\delta(x-y)+v_{0}\partial_x\delta(x-y),\\
\{v_0(x),v_2(y)\}&=&2v_{1,x}\delta(x-y)+2v_{1}\partial_x\delta(x-y),\\
\{v_1(x),v_2(y)\}&=&2v_{2,x}\delta(x-y)+3v_{2}\partial_x\delta(x-y)+h_1h_2(1+(N+1)(N-1)\alpha_0^2h_1h_2)\\
&&\cdot\left(\frac{1}{6}v_{0,xxx}\delta(x-y)+\frac{1}{2}v_{0,xx}\partial_x\delta(x-y)+\frac{1}{2}v_{0,x}\partial_x^2\delta(x-y)+\frac{1}{6}v_{0}\partial_x^3\delta(x-y)\right).
\eeaa
Define the pseudo-differential operator $L$
and the Hamiltonians
as before. The KP hierarchy can be described as the Poisson evolution equation
\be
\frac{\partial v_i(t,x)}{\partial t_n}=\{v_i,H_n\}.
\ee
To obtain the KP equation, we calculate
\bea
\frac{\partial v_0}{\partial t_3}&=&\frac {h_1h_2(1-{\alpha_0}^{2}h_1h_{2}\left( {N}^{2}+1 \right))}{2}v_{0,xxx}+3v_{2,x}+6v_0v_{0,x}-\frac{N}{h_1h_2}3v_{1,x},\label{v0t33J}\\
\frac{\partial v_1}{\partial t_2}&=&2v_{2,x}+\frac{h_1h_2(1+(N+1)(N-1)\alpha_0^2h_1h_2)}{6}v_{0,xxx}+2v_0v_{0,x},\label{v1t23J}\\
\frac{\partial v_0}{\partial t_2}&=&2v_{1,x}-\frac{N}{h_1h_2}2v_{0,x},\label{v0t23J}
\eea
From (\ref{v1t23J}) and (\ref{v0t23J}), we obtain
\beaa
v_1&=&\frac{1}{2}\partial_x^{-1}v_{0,t_2}+\frac{N}{h_1h_2}v_0,\\
v_2&=&\frac{1}{4}\partial_x^{-2}v_{0,t_2t_2}+\frac{N}{2h_1h_2}\partial_x^{-1}v_{0,t_2}-\frac{h_1h_2(1+(N+1)(N-1)\alpha_0^2h_1h_2)}{12}v_{0,xx}-\frac{1}{2}v_0^2,\\
\eeaa
substituting them into (\ref{v0t33J}), we obtain
\be
v_{0,t_3}=\frac{h_1h_2-\alpha_0^2h_1^2h_2^2(3N^2+1)}{4}v_{0,xxx}+\frac{3}{4}\partial_x^{-1}v_{0,t_2t_2}+3v_0v_{0,x}-\frac{N^2}{h_1^2h_2^2}3v_{0,x}.
\ee
Let
\be
\frac{\partial}{\partial t}:=\frac{\partial}{\partial t_3}+\frac{N^2}{h_1^2h_2^2}3\frac{\partial}{\partial x},
\ee
which corresponds to replacing the Hamiltonian $H_3$ by $H_3+\frac{N^2}{h_1^2h_2^2}3H_1$, then the equation above becomes
\[
v_{0,t}=\frac{h_1h_2-\alpha_0^2h_1^2h_2^2(3N^2+1)}{4}v_{0,xxx}+\frac{3}{4}\partial_x^{-1}v_{0,t_2t_2}+3v_0v_{0,x},
\]
Let $y=t_2$, then we have
\be
\frac{3}{4}v_{0,yy}=\left(v_{0,t}-\frac{h_1h_2-\alpha_0^2h_1^2h_2^2(3N^2+1)}{4}v_{0,xxx}-3v_0v_{0,x}\right)_x,
\ee
this is the KP equation. Clearly, it becomes the KP hierarchy in the last section when $N=1$.
From (\ref{tildev13J}), we see that
\[
v_0(x)=\sum_{m\in\Z}(a_{1,m}+ a_{2,m}+\cdots+a_{N,m})e^{-imx}.
\]
Let
\[
v_{j,0}(x)=\sum_{m\in\Z}a_{j,m}e^{-imx},
\]
which is associated to the slice on plane $z=j$ of 3D Young diagrams, then
\[
v_0=v_{1,0}+v_{2,0}+\cdots+v_{N,0}.
\]

Multiplied $v_0,x,y,t$ by a constant respectively and still denoted by $v_0,x,y,t$, the 3-KP equation becomes
\be
\frac{3}{4}v_{0,yy}=\left(v_{0,t}+\frac{h_1h_2-\alpha_0^2h_1^2h_2^2(3N^2+1)}{4}v_{0,xxx}-\frac{3}{2}v_0v_{0,x}\right)_x.\label{3-kpv0}
\ee
This equation is still called the KP equation. Let
$$u=-(h_1h_2-(3N^2+1)h_3^2)2(\text{log}\tau)_{xx},$$
 we get
\beaa
&&3\tau_{yy}\tau-3\tau_y\tau_y-4\tau_{tx}\tau+4\tau_t\tau_x\\
&&-\left(h_1h_2-(3N^2+1)h_3^2\right)\left(\tau_{xxxx}\tau-4\tau_{xxx}\tau_x+3\tau_{xx}\tau_{xx}\right)=0.
\eeaa
This equation of bilinear form can also be expressed by the 3-Jack polynomials, similar to (\ref{kpequationS}) and (\ref{kpequationJ}).

Note that the KP equation in this section is symmetric about $x$-axis and $y$-axis, that is, it is symmetric about $h_1$ and $h_2$, but $z$-axis is special, this is because we cut 3D Young diagrams into a series of 2D Young diagrams along $z$-axis. If we do that along $y$-axis, the results will be symmetric about $x$-axis and $z$-axis, and vice versa. Actually, the KP equation in this section is symmetric about $x$-axis, $y$-axis and $z$-axis.

In the following, we show the form of KP equation and the OPEs of the fields in $W_{1+\infty}$ algebra which are symmetric about $x,\ y,\ z$-axis, that is, they are symmetric about $h_1, h_2$ and $h_3$. From the representation of affine Yangian of $\mathfrak{gl}(1)$ on 3D Young diagrams, we see that the generators $\psi_j,\ e_j, f_j$ and the coefficients $\sigma_2,\ \sigma_3$ are symmetric about $h_1, h_2$ and $h_3$.

Let
\[
\psi_0=-\frac{N}{h_1h_2},
\]
we have
\beaa
&&k^2N+h_1h_2\alpha_0^2N(N+k)(N-k)\\
&=&k^3-(k+\psi_0h_1h_2)(k+\psi_0h_1h_3)(k+\psi_0h_2h_3)\nn\\
&=&-(k^2\psi_0\sigma_2+\psi_0^3\sigma_3^2)
\eeaa
for any number $k$. The OPE in (\ref{v1zv1wN}) becomes
\be
V_1(z)V_1(w)\sim \frac{\psi_0}{(z-w)^2}.
\ee
The OPE in (\ref{v2zv2wN}) becomes
\be
V_2(z)V_2(w)\sim \frac{-(\psi_0\sigma_2+\psi_0^3\sigma_3^2)}{2(z-w)^4}+\frac{2V_2{(w)}}{(z-w)^2}+\frac{V_2^\prime(w)}{z-w}.
\ee
The OPE in (\ref{v2zv3wN}) becomes
\be
V_2(z)V_3(w)\sim\frac{-(\sigma_2+\psi_0^2\sigma_3^2)V_1(w)}{(z-w)^4}+\frac{3V_3(w)}{(z-w)^2}+\frac{V_3^\prime(w)}{(z-w)}.
\ee
The OPE in (\ref{v1zv4wN}) becomes
\bea
V_1(z)V_4(w)&\sim &{\frac {-3\sigma_2+3\psi_0^2\sigma_3^2}{5}}\frac{V_1(w)}{(z-w)^4}+{\frac {3\sigma_2-3\psi_0^2\sigma_3^2}{5}}\frac{V_1'(w)}{(z-w)^3}\nn\\
&&+\frac{3V_3(w)}{(z-w)^2}+\frac{-\sigma_2+\psi_0^2\sigma_3^2}{10}\frac{V_1''(w)}{(z-w)^2}.
\eea
The OPE in (\ref{v2zv4wN}) becomes
\bea
V_2(z)V_4(w)&\sim&\frac{(21\sigma_2-9\psi_0^2\sigma_3^2)V_2(w)}{5(z-w)^4}+\frac{4V_4(w)}{(z-w)^2}+\frac{V_4'(w)}{(z-w)}.
\eea
Then by similar calculation, the KP equation (\ref{3-kpv0}) becomes
\be
\frac{3}{4}v_{0,yy}=\left(v_{0,t}+\frac{\sigma_2-3\psi_0^2\sigma_3^2}{4}v_{0,xxx}-\frac{3}{2}v_0v_{0,x}\right)_x.\label{3-kpv0Sy}
\ee
This equation is still called the KP equation. Let
$$u=-(\sigma_2-3\psi_0^2\sigma_3^2)2(\text{log}\tau)_{xx},$$
 we get
\bea
&&3\tau_{yy}\tau-3\tau_y\tau_y-4\tau_{tx}\tau+4\tau_t\tau_x\nn\\
&&-\left(\sigma_2-3\psi_0^2\sigma_3^2\right)\left(\tau_{xxxx}\tau-4\tau_{xxx}\tau_x+3\tau_{xx}\tau_{xx}\right)=0.\label{3-kpvoSyBili}
\eea
We can find that the KP equation in the form (\ref{3-kpv0Sy}) and (\ref{3-kpvoSyBili}) are symmetric about $h_1,\ h_2$ and $h_3$.
\section{The correspondence between Bosonic Fock space and Fermionic Fock space}\label{sect6}
In this section, we show the correspondence between the Bosonic Fock space and the Fermionic Fock space for three cases: the Schur functions, the Jack polynomials and the 3-Jack polynomials.
\subsection{For the Schur functions}
The Fermionic Fock representation space $\mathcal{F}$ is the space of Maya diagrams. A Maya diagram is made up of black and white stones lined up along the real line with the convention that all the stones are black far away to the right, whereas all the stones are white far away to the left. For example, the following is a Maya diagram\cite{MJD}
\begin{equation}\label{egmaya}
\begin{tikzpicture}
\draw (0,0) circle(0.1);
\draw (0,-0.5) node{$\frac{1}{2}$};
\fill (0.5,0) circle(0.1);
\draw (0.5,-0.5) node{$\frac{3}{2}$};
\draw (1,0) circle(0.1);
\draw (1,-0.5) node{$\frac{5}{2}$};
\fill (1.5,0) circle(0.1);
\draw (1.5,-0.5) node{$\frac{7}{2}$};
\fill (2,0) circle(0.1);
\draw (2,-0.5) node{$\frac{9}{2}$};
\draw (2.5,0) node{$\cdots$};
\fill (-0.5,0) circle(0.1);
\draw (-0.6,-0.5) node{$-\frac{1}{2}$};
\fill (-1,0) circle(0.1);
\draw (-1.1,-0.5) node{$-\frac{3}{2}$};
\draw (-1.5,0) circle(0.1);
\draw (-1.6,-0.5) node{$-\frac{5}{2}$};
\draw (-2,0) circle(0.1);
\draw (-2.1,-0.5) node{$-\frac{7}{2}$};
\draw (-2.5,0) node{$\cdots$};
\end{tikzpicture}
\end{equation}
By writing half integers $k_1, k_2,\cdots$ for the positions of the black stones,
a Maya diagram is described as an increasing sequence of half integers
\[
{\bf {k}}=\{k_n\}_{n\geq 1}\quad \text{with}\quad k_1<k_2<k_3<\cdots.
\]
For example, the Maya diagram in (\ref{egmaya}) is denoted by
\[
-\frac{3}{2},-\frac{1}{2},\frac{3}{2},\frac{7}{2},\frac{9}{2},\cdots
\]
Define the charge $p$ of a Maya diagram as the number of white stones on the right half line minus the number of black stones on the left half line. For example, the charge of Maya diagram in (\ref{egmaya}) is zero.
The basis vector is written as $|\mathbf{k}\rangle$. Specially,
\be\label{|0rangle}
|0\rangle=|\frac{1}{2},\frac{3}{2},\frac{5}{2},\cdots\rangle.
\ee
 The left action of the Fermions on $\mathcal{F}$ is defined as follows:
\begin{eqnarray}
 \psi_j|\mathbf{k}\rangle&=&\begin{cases}
  (-1)^{n-1}|\cdots,k_{n-1},k_{n+1},\cdots\rangle & \text{if}\,\,\,  k_n=-j;\\
   \quad\quad\quad 0 & \text{otherwise;}
\end{cases}\\
\psi^*_j|\mathbf{k}\rangle&=&\begin{cases}
 (-1)^{n}|\cdots,k_{n},j,k_{n+1},\cdots\rangle &\quad\text{if}\,\,\,  k_n<j<k_{n+1};\\
  \quad\quad \quad 0 & \quad\text{otherwise.}
  \end{cases}
\end{eqnarray}
Then Fermions $\psi^*_j,\ \psi_j,\ j\in\Z+\frac{1}{2}$ satisfy the anti-commutation relations
\begin{equation}\label{e:1.2}
\psi_j\psi_k+\psi_k\psi_j=0, \ \ \psi_j^*\psi_k^*+\psi_k^*\psi_j^*=0,\ \ \psi_j^*\psi_k+\psi_k\psi_j^*=\delta_{j+k,0}.
\end{equation}

Define\cite{3DFermionYangian,JackYangian}
 \be
\gamma_m=:\psi_{-(m+\frac{1}{2})}\psi_{m-\frac{1}{2}}^*:,\ \ \gamma_m^*=:\psi_{-m+\frac{1}{2}}\psi_{m+\frac{1}{2}}^*:.
\ee
Then they satisfy
\be
\gamma_i^2=0,\ \gamma_i^{*2}=0,\ \text{and}\ \gamma_{i}\gamma_{j}=\gamma_{j}\gamma_{i},\ \gamma_{i}^*\gamma_{j}^*=\gamma_{j}^*\gamma_{i}^*,\ \ \ \text{if}\ |i-j|>1.
\ee

It is known that it is isomorphic between the space of charged zero Maya diagrams and the space of 2D Young diagrams\cite{MJD}. Let a charged zero Maya diagram $|\bf{k}\rangle$ correspond to the 2D Young diagram $\lambda$. That the operator $\gamma_m$ (resp. $\gamma_m^*$) acting on $|\bf{k}\rangle$ corresponds to that the operator acting on $\lambda$ by adding (resp. removing) a box on the line $y-x=m$, which still denoted by $\gamma_m$ (resp. $\gamma_m^*$).
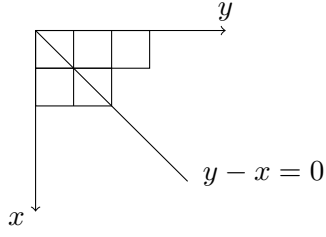
 \[
\begin{tikzpicture}
\draw (2.5,0.25) node{$y$};
\draw (0,0) -- (2.5,0) [->];
\draw (-0.25,-2.5) node{$x$};
\draw (0,0) -- (0,-2.4) [->];
\draw (0.5,-1) -- (0.5,0) [-];
\draw (1,-0.5) -- (1,0) [-];
\draw (0,0) -- (2,-2) [-];
\draw (0,-1)rectangle(1,-0.5);
\draw (0,-0.5)rectangle(1.5,0);
\draw (3,-1.9) node{$y-x=0$};
\end{tikzpicture}
\]

Define
\be
H_{n}=\sum_j[[[\gamma_{j}^*,\gamma_{j+1}^*],\cdots,\gamma_{j+n-1}^*],\gamma_{j+n}^*]
\ee
and
\be
H(\{p_n\})=\sum_{n=1}^\infty \frac{p_nH_n}{n}.
\ee
The Bosonic Fock space $\C[p_1,p_2,\cdots]$ is the space of Schur functions. We define the map $\sigma$ between the Fermionic Fock space and the Bosonic Fock space by
\be\label{sigmaSchur}
\sigma (|{\bf{k}}\rangle)=\langle 0|e^{H(\{p_n\})}|{\bf{k}}\rangle.
\ee
This map is the same with $\Phi$ in (5.15) in book \cite{MJD}. Therefore, this map gives the isomorphism between the Fermionic Fock space and the Bosonic Fock space. Since
\[
e^{H(p)}=\sum_\lambda S_\lambda(\{p_n\})S_\lambda^\bot(\{H_n\}),
\]
where $\bot$ means $S_\lambda^\bot(\{H_n\})$ is an annihilation operator.
let $|{\bf k}\rangle=|\lambda\rangle$ (the equality is the isomorphism between the space of charged zero Maya diagrams and the space of 2D Young diagrams),
\[
\sigma (|\lambda\rangle)=\langle 0|e^{H(\{p_n\})}|\lambda\rangle=S_{\lambda}(\{p_n\}),
\]
and
\[
\sigma (\gamma_m|\lambda\rangle)=\sigma (|\lambda+\Box_m\rangle)=S_{\lambda+\Box_m}(\{p_n\}),
\]
where $\lambda+\Box_m$ is a Young diagram obtained from $\lambda$ by adding a box on the line $y-x=m$ if this box is addable, zero if it is not addable.

\subsection{For the Jack polynomials}
We let $\psi_0=-\frac{1}{h_1h_2}$, where $\psi_0$ is a generator in the affine Yangian of $\mathfrak{gl}(1)$. Then in the representation space of the affine Yangian of $\mathfrak{gl}(1)$, the 3D Young diagrams which have more than one layer in $z$-axis direction become zero, therefore, the representation space of the affine Yangian of $\mathfrak{gl}(1)$ becomes the space of 2D Young diagrams in this special case.

Let the Fermionic Fock space be the space of Maya diagrams, and the Bosonic Space is the space of Jack polynomials $Y_\lambda$. We know that\cite{3DFermionYangian,JackYangian}
\[
f_0=-\sum_m \gamma_m^*,\ f_1=-\sum_m m\gamma_{m}^*.
\]
Define
\be
\mathcal{H}_n=-\frac{1}{(n-1)!}\text{ad}_{f_1}^{n-1}f_0,
\ee
and
\be
\mathcal{H}(\{p_n\})=\sum_{n=1}^\infty\frac{p_nH_n}{n\psi_0}.
\ee

Define the operator $Y_\lambda^\bot$ by
\be
\langle Y_\lambda Y_\mu, Y_\nu\rangle=\langle Y_\mu, Y_\lambda^\bot Y_\mu\rangle
\ee
as for Schur functions. Then for example, we have
\beaa
Y_\Box^\bot Y_{\begin{tikzpicture}
\draw [step=0.2](0,0) grid(.4,.2);
\end{tikzpicture}} &=&\psi_0\frac{\partial}{\partial{p_n}}\left(\frac{1}{h_1-h_2}p_{2}-\frac{h_2}{h_1-h_2}p_{1}^2\right)\\
&=&\frac{2}{h_1(h_1-h_2)}p_1=\frac{\langle Y_{\begin{tikzpicture}
\draw [step=0.2](0,0) grid(.4,.2);
\end{tikzpicture}},Y_{\begin{tikzpicture}
\draw [step=0.2](0,0) grid(.4,.2);
\end{tikzpicture}}\rangle}{\langle Y_\Box,Y_\Box\rangle}Y_\Box,
\eeaa
\beaa
Y_\Box^\bot Y_{\begin{tikzpicture}
\draw [step=0.2](0,0) grid(.4,.2);
\draw [step=0.2](0,-0.2) grid(.2,0);
\end{tikzpicture}_{h_1h_2}} &=&\psi_0\frac{\partial}{\partial{p_n}}\frac{1}{h_2-2h_1}\frac{1}{h_1-h_2}(2p_{3}-2(h_1+h_2)p_{1}p_{2}+2h_1h_2p_{1}^3)\\
&=&-\frac{1}{h_1h_2}\frac{1}{(h_2-2h_1)(h_1-h_2)}(6h_1h_2p_1^2-2(h_1+h_2)p_2)\\
&=&\frac{2(2h_2-h_1)}{h_2(h_2-2h_1)(h_2-h_1)}Y_{\begin{tikzpicture}
\draw [step=0.2](0,0) grid(.4,.2);
\end{tikzpicture}}+\frac{2}{h_1(h_1-h_2)}Y_{\begin{tikzpicture}
\draw [step=0.2](0,0) grid(.2,.4);
\end{tikzpicture}}\\
&=&\frac{\langle Y_{\begin{tikzpicture}
\draw [step=0.2](0,0) grid(.4,.2);
\draw [step=0.2](0,-0.2) grid(.2,0);
\end{tikzpicture}_{h_1h_2}},Y_{\begin{tikzpicture}
\draw [step=0.2](0,0) grid(.4,.2);
\draw [step=0.2](0,-0.2) grid(.2,0);
\end{tikzpicture}_{h_1h_2}}\rangle}{\langle Y_{\begin{tikzpicture}
\draw [step=0.2](0,0) grid(.4,.2);
\end{tikzpicture}},Y_{\begin{tikzpicture}
\draw [step=0.2](0,0) grid(.4,.2);
\end{tikzpicture}}\rangle}Y_{\begin{tikzpicture}
\draw [step=0.2](0,0) grid(.4,.2);
\end{tikzpicture}}+\frac{\varphi(h_2-h_1)\langle Y_{\begin{tikzpicture}
\draw [step=0.2](0,0) grid(.4,.2);
\draw [step=0.2](0,-0.2) grid(.2,0);
\end{tikzpicture}_{h_2h_1}},Y_{\begin{tikzpicture}
\draw [step=0.2](0,0) grid(.4,.2);
\draw [step=0.2](0,-0.2) grid(.2,0);
\end{tikzpicture}_{h_2h_1}}\rangle}{\langle Y_{\begin{tikzpicture}
\draw [step=0.2](0,0) grid(.2,.4);
\end{tikzpicture}},Y_{\begin{tikzpicture}
\draw [step=0.2](0,0) grid(.2,.4);
\end{tikzpicture}}\rangle}Y_{\begin{tikzpicture}
\draw [step=0.2](0,0) grid(.2,.4);
\end{tikzpicture}},
\eeaa
and
\beaa
Y_\Box^\bot Y_{\begin{tikzpicture}
\draw [step=0.2](0,0) grid(.4,.2);
\draw [step=0.2](0,-0.2) grid(.2,0);
\end{tikzpicture}_{h_2h_1}}
&=&\frac{\varphi(h_1-h_2)\langle Y_{\begin{tikzpicture}
\draw [step=0.2](0,0) grid(.4,.2);
\draw [step=0.2](0,-0.2) grid(.2,0);
\end{tikzpicture}_{h_1h_2}},Y_{\begin{tikzpicture}
\draw [step=0.2](0,0) grid(.4,.2);
\draw [step=0.2](0,-0.2) grid(.2,0);
\end{tikzpicture}_{h_1h_2}}\rangle}{\langle Y_{\begin{tikzpicture}
\draw [step=0.2](0,0) grid(.4,.2);
\end{tikzpicture}},Y_{\begin{tikzpicture}
\draw [step=0.2](0,0) grid(.4,.2);
\end{tikzpicture}}\rangle}Y_{\begin{tikzpicture}
\draw [step=0.2](0,0) grid(.4,.2);
\end{tikzpicture}}+\frac{\langle Y_{\begin{tikzpicture}
\draw [step=0.2](0,0) grid(.4,.2);
\draw [step=0.2](0,-0.2) grid(.2,0);
\end{tikzpicture}_{h_2h_1}},Y_{\begin{tikzpicture}
\draw [step=0.2](0,0) grid(.4,.2);
\draw [step=0.2](0,-0.2) grid(.2,0);
\end{tikzpicture}_{h_2h_1}}\rangle}{\langle Y_{\begin{tikzpicture}
\draw [step=0.2](0,0) grid(.2,.4);
\end{tikzpicture}},Y_{\begin{tikzpicture}
\draw [step=0.2](0,0) grid(.2,.4);
\end{tikzpicture}}\rangle}Y_{\begin{tikzpicture}
\draw [step=0.2](0,0) grid(.2,.4);
\end{tikzpicture}}.
\eeaa
Generally,
\[
Y_\lambda^\bot Y_\mu=\frac{\langle Y_\mu,Y_\mu\rangle}{\langle Y_{\mu/\lambda},Y_{\mu/\lambda}\rangle}Y_{\mu/\lambda}.
\]

We define the map $\sigma$ between the Fermionic Fock space (the space of Maya diagrams) and the Bosonic Fock space (the space of Jack polynomials $Y_\lambda$) by
\be\label{sigmaJack}
\sigma (|{\bf{k}}\rangle)=\langle 0|e^{\mathcal{H}(\{p_n\})}|{\bf{k}}\rangle.
\ee
This map sends the basis of the Fermionic Fock space to basis of the Bosonic Fock space. Since
\[
e^{\mathcal{H}(p)}=\sum_\lambda\frac{1}{\langle Y_\lambda,Y_\lambda\rangle} Y_\lambda(\{p_n\})Y_\lambda^\bot(\{\mathcal{H}_n\}).
\]
Let $|{\bf k}\rangle=|\lambda\rangle$ (the equality holds under the isomorphism between the space of charged zero Maya diagrams and the space of 2D Young diagrams),
\[
\sigma (|\lambda\rangle)=\langle 0|e^{\mathcal{H}(\{p_n\})}|\lambda\rangle=Y_{\lambda}(\{p_n\}),
\]
and
\[
\sigma (\gamma_m|\lambda\rangle)=\sigma (|\lambda+\Box_m\rangle)=Y_{\lambda+\Box_m}(\{p_n\}).
\]
Note that here $\langle {\bf k'}|{\bf k}\rangle$ does not equal $\delta_{{\bf k},{\bf k'}}$, but equal $\delta_{{\bf k},{\bf k'}}$ multiplied by $\langle Y_\lambda, Y_\lambda\rangle$, where $|{\bf k}\rangle=|\lambda\rangle$ under the isomorphism of the space of charged zero Maya diagrams and the space of 2D Young diagrams.

We see that when $h_1=1,h_2=-1,$ the results in this subsection become that in the last subsection.
\subsection{For 3-Jack polynomials}
We recall 3D Fermionic Fock space and 3D Bosonic Fock space. The 3D Bosonic Fock space $\C[\{P_{n,j\leq n}|n,j=1,2,\cdots\}]$ is the space of 3-Jack polynomials\cite{JHEP3Jack}. The elements in 3D Fermionic Fock space are the diagrams in the plane of the following form\cite{3DFermion}
\be
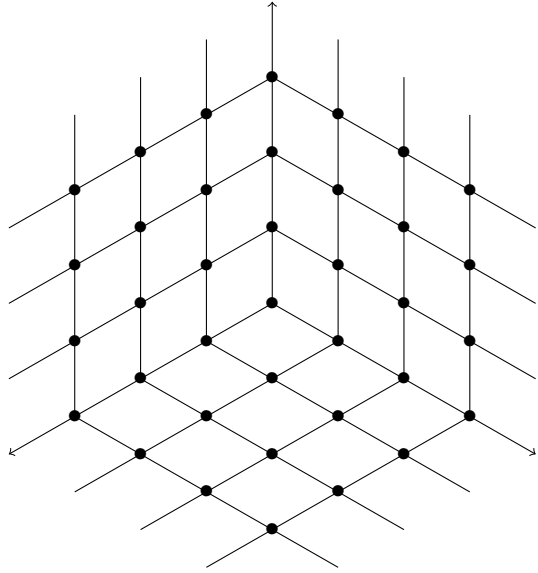
\begin{tikzpicture}
\draw (0,0) node{$\bullet$};
\draw (0,1) node{$\bullet$};
\draw (0,2) node{$\bullet$};
\draw (0,3) node{$\bullet$};
\draw (0,-1) node{$\bullet$};
\draw (0,-2) node{$\bullet$};
\draw (0,-3) node{$\bullet$};

\draw [shift = {+(0.866,-0.5)}](0,0) node{$\bullet$};
\draw [shift = {+(0.866,-0.5)}](0,1) node{$\bullet$};
\draw [shift = {+(0.866,-0.5)}](0,2) node{$\bullet$};
\draw [shift = {+(0.866,-0.5)}](0,3) node{$\bullet$};
\draw[shift = {+(0.866,-0.5)}] (0,-1) node{$\bullet$};
\draw[shift = {+(0.866,-0.5)}] (0,-2) node{$\bullet$};

\draw[shift = {+(0.866,-0.5)}] [shift = {+(0.866,-0.5)}](0,0) node{$\bullet$};
\draw[shift = {+(0.866,-0.5)}] [shift = {+(0.866,-0.5)}](0,1) node{$\bullet$};
\draw [shift = {+(0.866,-0.5)}][shift = {+(0.866,-0.5)}](0,2) node{$\bullet$};
\draw [shift = {+(0.866,-0.5)}][shift = {+(0.866,-0.5)}](0,3) node{$\bullet$};
\draw[shift = {+(0.866,-0.5)}][shift = {+(0.866,-0.5)}] (0,-1) node{$\bullet$};

\draw[shift = {+(0.866,-0.5)}][shift = {+(0.866,-0.5)}] [shift = {+(0.866,-0.5)}](0,0) node{$\bullet$};
\draw[shift = {+(0.866,-0.5)}][shift = {+(0.866,-0.5)}] [shift = {+(0.866,-0.5)}](0,1) node{$\bullet$};
\draw[shift = {+(0.866,-0.5)}] [shift = {+(0.866,-0.5)}][shift = {+(0.866,-0.5)}](0,2) node{$\bullet$};
\draw [shift = {+(0.866,-0.5)}][shift = {+(0.866,-0.5)}][shift = {+(0.866,-0.5)}](0,3) node{$\bullet$};

\draw [shift = {+(-0.866,-0.5)}](0,0) node{$\bullet$};
\draw [shift = {+(-0.866,-0.5)}](0,1) node{$\bullet$};
\draw [shift = {+(-0.866,-0.5)}](0,2) node{$\bullet$};
\draw [shift = {+(-0.866,-0.5)}](0,3) node{$\bullet$};
\draw[shift = {+(-0.866,-0.5)}] (0,-1) node{$\bullet$};
\draw[shift = {+(-0.866,-0.5)}] (0,-2) node{$\bullet$};

\draw[shift = {+(-0.866,-0.5)}] [shift = {+(-0.866,-0.5)}](0,0) node{$\bullet$};
\draw[shift = {+(-0.866,-0.5)}] [shift = {+(-0.866,-0.5)}](0,1) node{$\bullet$};
\draw [shift = {+(-0.866,-0.5)}][shift = {+(-0.866,-0.5)}](0,2) node{$\bullet$};
\draw [shift = {+(-0.866,-0.5)}][shift = {+(-0.866,-0.5)}](0,3) node{$\bullet$};
\draw[shift = {+(-0.866,-0.5)}][shift = {+(-0.866,-0.5)}] (0,-1) node{$\bullet$};

\draw[shift = {+(-0.866,-0.5)}][shift = {+(-0.866,-0.5)}] [shift = {+(-0.866,-0.5)}](0,0) node{$\bullet$};
\draw[shift = {+(-0.866,-0.5)}][shift = {+(-0.866,-0.5)}] [shift = {+(-0.866,-0.5)}](0,1) node{$\bullet$};
\draw[shift = {+(-0.866,-0.5)}] [shift = {+(-0.866,-0.5)}][shift = {+(-0.866,-0.5)}](0,2) node{$\bullet$};
\draw [shift = {+(-0.866,-0.5)}][shift = {+(-0.866,-0.5)}][shift = {+(-0.866,-0.5)}](0,3) node{$\bullet$};

\draw (0,0) -- (0,4) [->];
\draw [shift = {+(0.866,-0.5)}](0,0) -- (0,4) [-];
\draw[shift = {+(0.866,-0.5)}][shift = {+(0.866,-0.5)}] (0,0) -- (0,4) [-];
\draw [shift = {+(0.866,-0.5)}][shift = {+(0.866,-0.5)}][shift = {+(0.866,-0.5)}](0,0) -- (0,4) [-];
\draw[shift = {+(-0.866,-0.5)}] (0,0) -- (0,4) [-];
\draw [shift = {+(-0.866,-0.5)}][shift = {+(-0.866,-0.5)}](0,0) -- (0,4) [-];
\draw [shift = {+(-0.866,-0.5)}][shift = {+(-0.866,-0.5)}][shift = {+(-0.866,-0.5)}](0,0) -- (0,4) [-];

\draw (0,0) -- (3.464,-2) [->];
\draw[shift = {+(0,1)}] (0,0) -- (3.464,-2) [-];
\draw [shift = {+(0,1)}][shift = {+(0,1)}](0,0) -- (3.464,-2) [-];
\draw[shift = {+(0,1)}][shift = {+(0,1)}][shift = {+(0,1)}] (0,0) -- (3.464,-2) [-];
\draw [shift = {+(-0.866,-0.5)}](0,0) -- (3.464,-2) [-];
\draw [shift = {+(-0.866,-0.5)}][shift = {+(-0.866,-0.5)}](0,0) -- (3.464,-2) [-];
\draw [shift = {+(-0.866,-0.5)}][shift = {+(-0.866,-0.5)}][shift = {+(-0.866,-0.5)}](0,0) -- (3.464,-2) [-];

\draw (0,0) -- (-3.464,-2) [->];
\draw[shift = {+(0,1)}] (0,0) -- (-3.464,-2) [-];
\draw [shift = {+(0,1)}][shift = {+(0,1)}](0,0) -- (-3.464,-2) [-];
\draw[shift = {+(0,1)}][shift = {+(0,1)}][shift = {+(0,1)}] (0,0) -- (-3.464,-2) [-];
\draw [shift = {+(0.866,-0.5)}](0,0) -- (-3.464,-2) [-];
\draw [shift = {+(0.866,-0.5)}][shift = {+(0.866,-0.5)}](0,0) -- (-3.464,-2) [-];
\draw [shift = {+(0.866,-0.5)}][shift = {+(0.866,-0.5)}][shift = {+(0.866,-0.5)}](0,0) --(-3.464,-2) [-];
\end{tikzpicture}
\ee
This state is the vacuum state denoted by $|0\rangle$. The black point is called a vertex.  At every vertex $\overrightarrow{m}$, there are six half-lines connecting it with its neighbors:
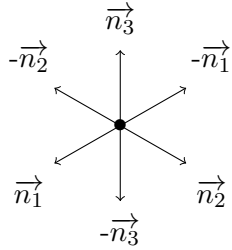
\be
\begin{tikzpicture}
\draw (0,0) node{$\bullet$};

\draw (0,0) -- (0.866,-0.5) [->];
\draw (1.2,-0.9) node{$\overrightarrow{n_2}$};

\draw (0,0) -- (0.866,0.5) [->];
\draw (1.2,0.9) node{-$\overrightarrow{n_1}$};

\draw (0,0) -- (-0.866,-0.5) [->];
\draw (-1.2,-0.9) node{$\overrightarrow{n_1}$};

\draw (0,0) -- (-0.866,0.5) [->];
\draw (-1.2,0.9) node{-$\overrightarrow{n_2}$};

\draw (0,0) -- (0,1) [->];
\draw (0,1.4) node{$\overrightarrow{n_3}$};

\draw (0,0) -- (0,-1) [->];
\draw (0,-1.4) node{-$\overrightarrow{n_3}$};
\end{tikzpicture}
\ee
where $\overrightarrow{n_1}=(-\frac{\sqrt 3}{2},-\frac{1}{2})$, $\overrightarrow{n_2}=(\frac{\sqrt 3}{2},-\frac{1}{2})$, $\overrightarrow{n_3}=(0,1)$.
A state at $\overrightarrow{m}$ is characterized by some of the arrows above. For example, the state at $\overrightarrow{m}$ can be
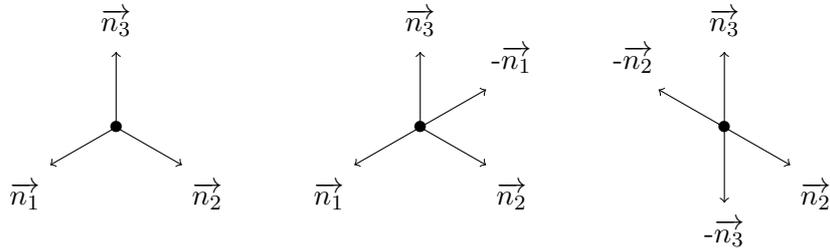
\be
\begin{tikzpicture}
\draw(0,0) node{$\bullet$};

\draw (0,0) -- (0.866,-0.5) [->];
\draw (1.2,-0.9) node{$\overrightarrow{n_2}$};

\draw (0,0) -- (-0.866,-0.5) [->];
\draw (-1.2,-0.9) node{$\overrightarrow{n_1}$};

\draw (0,0) -- (0,1) [->];
\draw (0,1.4) node{$\overrightarrow{n_3}$};

\draw[shift = {+(4,0)}] (0,0) node{$\bullet$};

\draw [shift = {+(4,0)}](0,0) -- (0.866,-0.5) [->];
\draw [shift = {+(4,0)}](1.2,-0.9) node{$\overrightarrow{n_2}$};

\draw [shift = {+(4,0)}](0,0) -- (0.866,0.5) [->];
\draw[shift = {+(4,0)}] (1.2,0.9) node{-$\overrightarrow{n_1}$};

\draw[shift = {+(4,0)}] (0,0) -- (-0.866,-0.5) [->];
\draw[shift = {+(4,0)}] (-1.2,-0.9) node{$\overrightarrow{n_1}$};

\draw [shift = {+(4,0)}](0,0) -- (0,1) [->];
\draw[shift = {+(4,0)}] (0,1.4) node{$\overrightarrow{n_3}$};

\draw[shift = {+(8,0)}] (0,0) node{$\bullet$};

\draw [shift = {+(8,0)}](0,0) -- (0.866,-0.5) [->];
\draw[shift = {+(8,0)}] (1.2,-0.9) node{$\overrightarrow{n_2}$};

\draw [shift = {+(8,0)}](0,0) -- (-0.866,0.5) [->];
\draw[shift = {+(8,0)}] (-1.2,0.9) node{-$\overrightarrow{n_2}$};

\draw[shift = {+(8,0)}] (0,0) -- (0,1) [->];
\draw [shift = {+(8,0)}](0,1.4) node{$\overrightarrow{n_3}$};

\draw [shift = {+(8,0)}](0,0) -- (0,-1) [->];
\draw[shift = {+(8,0)}] (0,-1.4) node{-$\overrightarrow{n_3}$};
\end{tikzpicture}
\ee
and so on. Other elements in the 3D Fermionic Fock space can be obtained from the vacuum state by the actions of 3D Fermions. 3D Fermions $\Gamma_{\overrightarrow{m}}$ and $\Gamma_{\overrightarrow{m}}^*$ (corresponding to $\gamma_m$ and $\gamma_m^*$ for 2D Young diagrams respectively) act on states by
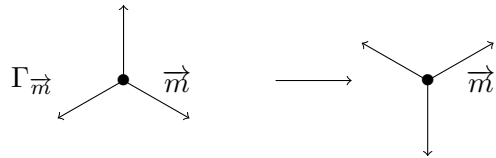
\be\label{gammaaction}
\begin{tikzpicture}
\draw (-1.2,0) node{$\Gamma_{\overrightarrow{m}}$};
\draw (0,0) node{$\bullet$};
\draw (0.7,0) node{$\overrightarrow{m}$};
\draw (4.7,0) node{$\overrightarrow{m}$};

\draw (0,0) -- (0.866,-0.5) [->];

\draw (2,0) -- (3,0) [->];

\draw (0,0) -- (-0.866,-0.5) [->];

\draw (0,0) -- (0,1) [->];

\draw[shift = {+(4,0)}] (0,0) node{$\bullet$};

\draw [shift = {+(4,0)}](0,0) -- (-0.866,0.5) [->];

\draw[shift = {+(4,0)}] (0,0) -- (0.866,0.5) [->];

\draw [shift = {+(4,0)}](0,0) -- (0,-1) [->];

\end{tikzpicture}
\ee
and
\be\label{gamma*action}
\begin{tikzpicture}
\draw (-1.2,0) node{$\Gamma_{\overrightarrow{m}}^*$};
\draw (0,0) node{$\bullet$};
\draw (0.7,0) node{$\overrightarrow{m}$};
\draw (4.7,0) node{$\overrightarrow{m}$};

\draw [shift = {+(4,0)}](0,0) -- (0.866,-0.5) [->];

\draw (2,0) -- (3,0) [->];

\draw [shift = {+(4,0)}](0,0) -- (-0.866,-0.5) [->];

\draw [shift = {+(4,0)}](0,0) -- (0,1) [->];

\draw[shift = {+(4,0)}] (0,0) node{$\bullet$};

\draw (0,0) -- (-0.866,0.5) [->];

\draw (0,0) -- (0.866,0.5) [->];

\draw(0,0) -- (0,-1) [->];

\end{tikzpicture}
\ee
They send other states to zero.
Then it is clear that the 3D Fermionic Fock space is isomorphic to the space of 3D Young diagrams. We also use $\pi$ to denote the elements in 3D Fermionic Fock space.

From \cite{3DFermionYangian}, we know that the generators $e_j,\ f_j,\ \psi_j$ can be represented by $\Gamma_{\overrightarrow{m}}$ and $\Gamma_{\overrightarrow{m}}^*$:
\[
e_j=\sum_{\overrightarrow{m}}m^j\Gamma_{\overrightarrow{m}},\ \ f_j=-\sum_{\overrightarrow{m}}m^j\Gamma_{\overrightarrow{m}}^*,\ \ \psi_j=\sum_{\overrightarrow{m}}m^j[\Gamma_{\overrightarrow{m}}^*,\Gamma_{\overrightarrow{m}}].
\]
In \cite{JHEP3DBoson}, we construct the 3D Bosons $b_{n,j}$ and show the relations between $b_{n,j}$ and the generators $V_{j,n}$ of $W_{1+\infty}$ algebra, and we found that the variables $P_{n,j}=b_{n,j}|0\rangle.$ The affine Yangian of $\mathfrak{gl}(1)$ is isomorphic to the universal enveloping algebra of $W_{1+\infty}$\cite{GGL}. Then 3D Bosons $b_{n,j}$ can be represented by $\Gamma_{\overrightarrow{m}}$ and $\Gamma_{\overrightarrow{m}}^*$, for example,
\beaa
b_{-n,1}&=&\frac{1}{\left(n-1\right)!}\text{ad}_{\left(\sum_{\overrightarrow{m}}m\Gamma_{\overrightarrow{m}}\right)}^{n-1}\left(\sum_{\overrightarrow{m}}\Gamma_{\overrightarrow{m}}\right),\\
b_{n,1}&=&-\frac{1}{\left(n-1\right)!}\text{ad}_{\left(-\sum_{\overrightarrow{m}}m\Gamma_{\overrightarrow{m}}^*\right)}^{n-1}\left(-\sum_{\overrightarrow{m}}m^j\Gamma_{\overrightarrow{m}}^*\right)\\
\eeaa
and
\beaa
b_{-\left(n+1\right),2}&=&\frac{1}{\left(n-1\right)!}\text{ad}_{\left(\sum_{\overrightarrow{m}}m\Gamma_{\overrightarrow{m}}\right)}^{n-1}
\left([\left(\sum_{\overrightarrow{m}}m^2\Gamma_{\overrightarrow{m}}\right),\left(\sum_{\overrightarrow{m}}\Gamma_{\overrightarrow{m}}\right)]\right.\\
&&\left.-\sigma_3[\left(\sum_{\overrightarrow{m}}m\Gamma_{\overrightarrow{m}}\right),\left(\sum_{\overrightarrow{m}}\Gamma_{\overrightarrow{m}}\right)]\right)-\sum_{i+j=-\left(n+1\right)}:b_{i,1}b_{j,1}:,\\
b_{n+1,2}&=&-\frac{1}{\left(n-1\right)!}\text{ad}_{\left(-\sum_{\overrightarrow{m}}m\Gamma_{\overrightarrow{m}}^*\right)}^{n-1}\left([\left(\sum_{\overrightarrow{m}}m^2\Gamma_{\overrightarrow{m}}^*\right),\left(-\sum_{\overrightarrow{m}}\Gamma_{\overrightarrow{m}}^*\right)]\right.\\
&&\left.-\sigma_3[\left(-\sum_{\overrightarrow{m}}m\Gamma_{\overrightarrow{m}}^*\right),\left(-\sum_{\overrightarrow{m}}\Gamma_{\overrightarrow{m}}^*\right)]\right)-\sum_{i+j=\left(n+1\right)}:b_{i,1}b_{j,1}:
\eeaa
for $n\geq 1$.
This is the realization of 3D Bosons by 3D Fermions. When $\psi_0=-\frac{1}{h_1h_2}$, this realization becomes that for the deformed KP hierarchy corresponding to that the space of 3D Young diagrams becomes the space of 2D Young diagrams. When $h_1=1,h_2=-1, \psi_0=1$, this realization becomes $H_n=\sum_{j\in\Z+1/2}:\psi_{-j}\psi_{j+n}^*:$ for the KP hierarchy. Define
\be
\mathcal{H}^{3D}(\{P_{n,j\leq n}\})=\sum_{n\geq 1,j}\frac{P_{n,j}b_{n,j}}{\langle P_{n,j},P_{n,j}\rangle}.
\ee
For any 3D Young diagrams $\pi$, the 3-Jack polynomials are denoted by $J_{\pi}$, we define the annihilation operator $J_\pi^\bot$ by
\[
\langle J_{\pi_1}J_{\pi_2},J_{\pi_3}\rangle=\langle J_{\pi_2}, J_{\pi_1}^\bot J_{\pi_3}\rangle.
\]
Then
\be\label{H3DJack}
e^{\mathcal{H}^{3D}(\{P_{n,j\leq n}\})}=\sum_{\pi}\frac{1}{\langle J_\pi,J_\pi\rangle} J_{\pi}(\{P_{n,j\leq n}\})J_\pi^\bot(\{b_{n,j\leq n}\}).
\ee
Define the map $\sigma$ from the 3D Fermionic Fock space to the 3D Bosonic Fock space by
\be
\sigma(|\pi\rangle)=\langle 0|e^{\mathcal{H}^{3D}(\{P_{n,j\leq n}\})}|\pi\rangle.
\ee
From (\ref{H3DJack}), we get
\[
\sigma(|\pi\rangle)=J_\pi.
\]
\section*{Appendix A: The calculations in OPEs $V_j(z)V_4(w)$}
Here, we list our calculations to obtain the OPEs $V_j(z)V_4(w), j=1,2$. From
\[
U_1(z)U_1(w)\sim -\frac{N}{h_1h_2}\frac{1}{(z-w)^2},
\]
we obtain
\beaa
V_1(z)U_1'U_1U_1(w)&\sim&-\frac{2N}{h_1h_2}\frac{U_1U_1(w)}{(z-w)^3}-\frac{2N}{h_1h_2}\frac{U_1'U_1(w)}{(z-w)^2},\\
V_1(z)U_1'U_1'(w)&\sim&-\frac{4N}{h_1h_2}\frac{U_1'(w)}{(z-w)^3},\\
V_1(z)U_1U_1U_1U_1(w)&\sim&-\frac{4N}{h_1h_2}\frac{U_1U_1U_1(w)}{(z-w)^2},\\
V_1(z)U_1''U_1(w)&\sim&-\frac{6N}{h_1h_2}\frac{U_1(w)}{(z-w)^4}-\frac{N}{h_1h_2}\frac{U_1''(w)}{(z-w)^2},\\
V_1(z)U_1'''(w)&\sim&-\frac{24N}{h_1h_2}\frac{1}{(z-w)^5}.
\eeaa
From
\beaa
U_1(z)U_2(w)&\sim& -\frac{N(N-1)\alpha_0}{h_1h_2}\frac{1}{(z-w)^3}-\frac{(N-1)}{h_1h_2}\frac{U_1(w)}{(z-w)^2},\\
U_2(z)U_2(w)&\sim& \frac{N(N-1)(1+2(2N-1)\alpha_0h_1h_2)}{2h_1^2h_2^2(z-w)^4}+\frac{2U_2(w)}{h_1h_2(z-w)^2}\\
&&-\frac{(N-1)U_1U_1(w)}{h_1h_2(z-w)^2}-\frac{N(N-1)\alpha_0U_1'(w)}{h_1h_2(z-w)^2}\\
&&+\frac{U_2'(w)}{h_1h_2(z-w)}-\frac{(N-1)U_1'U_1(w)}{h_1h_2(z-w)}-\frac{N(N-1)\alpha_0U_1''(w)}{2h_1h_2(z-w)},
\eeaa
we obtain
\beaa
V_1(z)U_1'U_2(w)&\sim&{-\frac {2N{ U_2(w)}}{h_{1}h_{2} ( z-w ) ^{3}}}-{
\frac { \left( N-1 \right) N\alpha_0{ U_1'(w)}}{h_{1}h_{2} \left( z-w
 \right) ^{3}}}-{\frac { \left( N-1 \right) { U_1' U_1(w)}}{h_{1}
h_{2} \left( z-w \right) ^{2}}},\\
V_1(z)U_1U_2'(w)&\sim&-{\frac { 3\left( N-1 \right) N\alpha_0{ U_1(w)}}{h_{1}h_{2} \left( z-w
 \right) ^{4}}}-{\frac { 2\left( N-1 \right) {{ U_1U_1(w)}}}{h_{1}
h_{2} \left( z-w \right) ^{3}}}\\&&-{\frac { \left( N-1 \right) { U_1'(w)
}{ U_1(w)}}{h_{1}h_{2} \left( z-w \right) ^{2}}}-{\frac {N{
U_2'(w)}}{h_{1}h_{2}\left( z-w \right) ^{2}}},\\
V_1(z)U_1U_1U_2(w)&\sim&-{\frac {N \left( N-1 \right) \alpha_0{{ U_1U_1(w)}}}{h_{1}h_{2} \left( z
-w \right) ^{3}}}-2{\frac {N{ U_1 U_2(w)}}{h_{1}h_{2}
 \left( z-w \right) ^{2}}}-{\frac { \left( N-1 \right) {{ U_1U_1U_1(w)}}
}{h_{1}h_{2} \left( z-w \right) ^{2}}}
,\\
V_1(z)U_2U_2(w)&\sim&4{\frac {N \left( N-1 \right) ^{2}\alpha_0}{{h_{1}}^{2}{h_{2}}^{2} \left( z
-w \right) ^{5}}}+3{\frac { \left( N-1 \right) ^{2}{ U_1(w)}}{{h_{1}}
^{2}{h_{2}}^{2} \left( z-w \right) ^{4}}}-2{\frac { \left( N-1
 \right) N\alpha_0{ U_2(w)}}{h_{1}h_{2} \left( z-w \right) ^{3}}}\\&&-2{
\frac { \left( N-1 \right) {U_1 U_2(w)}}{h_{1}h_{2}\left( z
-w \right) ^{2}}}+{\frac { \left( N-1 \right) ^{2}{ U_1''(w)}}{2{h_
{1}}^{2}{h_{2}}^{2} \left( z-w \right) ^{2}}}
,\\
V_1(z)U_2''(w)&\sim&-12{\frac { \left( N-1 \right) N\alpha_0}{h_{1}h_{2} \left( z-w
 \right) ^{5}}}-6{\frac { \left( N-1 \right) { U_1(w)}}{h_{1}h_{2}
 \left( z-w \right) ^{4}}}-4{\frac { \left( N-1 \right) { U_1'(w)}
}{h_{1}h_{2} \left( z-w \right) ^{3}}}-{\frac { \left( N-1
 \right) {U_1''(w)}}{h_{1}h_{2} \left( z-w \right) ^{2}}}
.
\eeaa

From
\beaa
U_1(z)U_3(w)\sim -\frac{N(N-1)(N-2)\alpha_0^2}{h_1h_2(z-w)^4}-\frac{(N-1)(N-2)\alpha_0U_1(w)}{h_1h_2(z-w)^3}
-\frac{(N-2)U_2(w)}{h_1h_2(z-w)^2},
\eeaa
we obtain
\beaa
V_1(z)U_3'(w)&\sim&-4{\frac {N \left( N-1 \right)  \left( N-2 \right) {\alpha_0}^{2}}{h_{1}h
_{2} \left( z-w \right) ^{5}}}-3{\frac { \left( N-1 \right)
 \left( N-2 \right) \alpha_0{ U_1(w)}}{h_{1}h_{2} \left( z-w \right) ^{4}}
}\\
&&-2{\frac { \left( N-2 \right) { U_2(w)}}{h_{1}h_{2} \left( z-w
 \right) ^{3}}}-{\frac { \left( N-1 \right)  \left( N-2 \right) \alpha_0{
U_1'(w)}}{h_{1}h_{2}\left( z-w \right) ^{3}}}-{\frac { \left( N-2
 \right) { U_2'(w)}}{h_{1}h_{2} \left( z-w \right) ^{2}}},\\
V_1(z)U_1U_3(w)&\sim&-{\frac {N \left( N-1 \right)  \left( N-2 \right) {\alpha_0}^{2}{ U_1(w)}}{h_{
1}h_{2} \left( z-w \right) ^{4}}}-{\frac { \left( N-1 \right)
 \left( N-2 \right) \alpha_0{{ U_1(w)}}^{2}}{h_{1}h_{2} \left( z-w
 \right) ^{3}}}\\&&-{\frac {N{ U_3(w)}}{h_{1}h_{2} \left( z-w \right) ^
{2}}}-{\frac { \left( N-2 \right) { U_1U_2(w)}}{h_{1}h_{2} \left( z-w
 \right) ^{2}}}.
\eeaa
We also have
\beaa
V_1(z)U_4(w)&\sim & -\frac{N(N-1)(N-2)(N-3)\alpha_0^3}{h_1h_2(z-w)^5}-\frac{(N-1)(N-2)(N-3)\alpha_0^2U_1(w)}{h_1h_2(z-w)^4}\\&&
-\frac{(N-2)(N-3)\alpha_0U_2(w)}{h_1h_2(z-w)^3}
-\frac{(N-3)U_3(w)}{h_1h_2(z-w)^2}.
\eeaa

From
\[
V_2(z)U_1(w)\sim \frac{U_1(w)}{(z-w)^2}+\frac{U_1^\prime(w)}{(z-w)},
\]
we have
\beaa
V_2(z)U_1'U_1U_1(w)&\sim&{-\frac {4N{ U_1(w)}}{h_{1}h_{2} ( z-w ) ^{5}}}-{
\frac {   N{ U_1'(w)}}{h_{1}h_{2} \left( z-w
 \right) ^{4}}}+{\frac { 2 { U_1U_1 U_1(w)}}{ \left( z-w \right) ^{3}}}\\
 &&+\frac{4U_1'U_1U_1(w)}{(z-w)^2}+\frac{2U_1'U_1'U_1(w)+U_1''U_1U_1(w)}{z-w},\\
V_2(z)U_1'U_1'(w)&\sim&-\frac{4N}{h_1h_2(z-w)^6}+\frac{4 U_1' U_1(w)}{(z-w)^3}+\frac{4 U_1'U_1'(w)}{(z-w)^2}+\frac{2 U_1'' U_ 1'(w)}{z-w},\\
V_2(z)U_1U_1U_1U_1(w)&\sim&-\frac{6 N U_1U_1(w)}{h_1 h_2(z-w)^4}+\frac{4 U_1U_1U_1U_1(w)}{(z-w)^2}+\frac{4 U_1' U_1U_1U_1(w)}{z-w},\\
V_2(z)U_1''U_1(w)&\sim&-\frac{6 N}{h_1 h_2(z-w)^6}+\frac{6 U_1U_1(w)}{(z-w)^4}+\frac{6 U_1' U_1(w)}{(z-w)^3}+\frac{4 U_1'' U_1(w)}{(z-w)^2}\\
&&+\frac{U_1''' U_1(w)+U_1'' U_1'(w)}{z-w},\\
V_2(z)U_1'''(w)&\sim&\frac{24 U_1(w)}{(z-w)^5}+\frac{24 U_1'(w)}{(z-w)^4}+\frac{12 U_1''(w)}{(z-w)^3}+\frac{4 U_1'''(w)}{(z-w)^2}+\frac{U_1^{(4)}(w)}{z-w}.
\eeaa

From
\beaa
V_2(z)U_2(w)&\sim& \frac{(N^3-N)\alpha_0^2}{2(z-w)^4}+\frac{(N-1)\alpha_0U_1(w)}{(z-w)^3}+\frac{2U_2(w)}{(z-w)^2}+\frac{U_2'(w)}{(z-w)},\\
\eeaa
we obtain
\beaa
V_2(z)U_1'U_2(w)&\sim&-\frac{2 N(N-1) \alpha_0}{h_1 h_2(z-w)^6}-\frac{2(N-1) U_1(w)}{h_1 h_2(z-w)^5}+\frac{\left(N^3-N\right) \alpha_0^2 U_1'(w)}{2(z-w)^4}\\
&&+\frac{2 U_1 U_2(w)+\alpha_0(N-1) U_1' U_1(w)}{(z-w)^3}\\
&&+\frac{4 U_1' U_2(w)}{(z-w)^2}+\frac{U_1' U_2'(w)+U_1'' U_2(w)}{z-w},\\
V_2(z)U_1U_2'(w)&\sim&-\frac{3 N(N-1) \alpha_0}{h_1 h_2(z-w)^6}+\frac{\left(2\left(N^3-N\right) \alpha_0^2-\frac{2(N-1)}{h_1 h_2}\right) U_1(w)}{(z-w)^5}\\
&&+\frac{3 \alpha_0(N-1) U_1U_1(w)-\frac{(N-1) U_1'(w)}{h_1 h_2}}{(z-w)^4} \\
&&+\frac{4 U_1 U_2(w)+\alpha_0(N-1) U_1' U_1(w)}{(z-w)^3}+\frac{4 U_1 U_2'(w)}{(z-w)^2}+\frac{U_1 U_2''(w)+U_1' U_2'(w)}{z-w},\\
V_2(z)U_1U_1U_2(w)&\sim&-\frac{2 \alpha_0 N(N-1) U_1(w)}{h_1 h_2(z-w)^5}+\frac{\left(\left(N^3-N\right) \alpha_0^2-\frac{4(N-1)}{h_1 h_2}\right) U_1U_1(w)-\frac{2 N U_2(w)}{h_1 h_2}}{2(z-w)^4}\\
&&+\frac{\alpha_0(N-1) U_1U_1U_1(w)}{(z-w)^3}+\frac{4 U_1U_1 U_2(w)}{(z-w)^2}+\frac{U_1U_1 U_2(w)+2 U_1' U_1 U_2(w)}{z-w}
,\\
V_2(z)U_2U_2(w)&\sim&\frac{\frac{3 N(N-1)}{h_1^2 h_2^2}+\frac{2 \alpha_0^2 N(N-1)(N+2)}{h_1 h_2}}{(z-w)^6}-\frac{6 \alpha_0(N-1)^2 U_1(w)}{h_1 h_2(z-w)^5}\\
&&+\frac{\left(\left(N^3-N\right) \alpha_0^2+\frac{8}{h_1 h_2}\right) U_2(w)-\frac{4 \alpha_0 N(N-1) U_1'(w)}{h_1 h_2}-\frac{4(N-1) U_1U_1(w)}{h_1 h_2}}{(z-w)^4}\\
&&+\frac{2 \alpha_0(N-1) U_1 U_2(w)-\frac{3(N-1) U_1' U_1(w)}{h_1 h_2}+\frac{3 U_2'(w)}{h_1 h_2}-\frac{\alpha_0(N-1)(4 N-1) U_1''}{2 h_1 h_2}}{(z-w)^3} +\frac{4 U_2U_2(w)}{(z-w)^2}\\
&&+\frac{2 U_2' U_2(w)-\frac{(N-1) U_1'' U_1'(w)}{2 h_1 h_2}-\frac{(N-1) U_1''' U_1(w)}{6 h_1 h_2}+\frac{U_2'''(w)}{6 h_1 h_2}-\frac{\alpha_0 N(N-1) U_1^{(4)}(w)}{12 h_1 h_2}}{z-w}
,\\
V_2(z)U_2''(w)&\sim&\frac{10 \alpha_0^2\left(N^3-N\right)}{(z-w)^6}+\frac{12 \alpha_0(N-1) U_1(w)}{(z-w)^5}+\frac{12 U_2(w)+6 \alpha_0(N-1) U_1'(w)}{(z-w)^4}\\
&&+\frac{10 U_2'(w)+\alpha_0(N-1) U_1''(w)}{(z-w)^3} +\frac{4 U_2''(w)}{(z-w)^2}+\frac{U_2'''(w)}{z-w}
.
\eeaa
From
\beaa
V_2(z)U_3(w)&\sim&\frac{(N+1)N(N-1)(N-2)\alpha_0^3}{(z-w)^5}
+\frac{(N+3)(N-1)(N-2)\alpha_0^2U_1(w)}{2(z-w)^4}\\
&&+\frac{2(N-2)\alpha_0U_2(w)}{(z-w)^3}+
\frac{3U_3(w)}{(z-w)^2}+\frac{U_3'(w)}{(z-w)},
\eeaa
we get
\beaa
V_2(z)U_3'(w)&\sim&\frac{5 \alpha_0^3(N+1) N(N-1)(N-2)}{(z-w)^6}+\frac{2 \alpha_0^2(N+3)(N-1)(N-2) U_1(w)}{(z-w)^5} \\
&&+\frac{6(N-2) \alpha_0 U_2(w)+\frac{\alpha_0^2(N+3)(N-1)(N-2) U_1'(w)}{2}}{(z-w)^4}+\frac{6 U_3(w)+2 a(N-2) U_2'(w)}{(z-w)^3}\\
&&+\frac{4 U_3'(w)}{(z-w)^2}+\frac{U_3''(w)}{z-w},\\
V_2(z)U_1U_3(w)&\sim&-\frac{N(N-1)(N-2) \alpha_0^2}{h_1 h_2(z-w)^6}+\frac{\left(\alpha_0^3(N+1) N(N-1)(N-2)-\frac{(N-1)(N-2) \alpha_0}{h_1 h_2}\right) U_1(w)}{(z-w)^5}\\
&& +\frac{\frac{\alpha_0^2(N+3)(N-1)(N-2) U_1U_1(w)}{2}-\frac{(N-2) U_2(w)}{h_1 h_2}}{(z-w)^4}+\frac{2 \alpha_0(N-2) U_1 U_2(w)}{(z-w)^3}\\
&&+\frac{4 U_1 U_3(w)}{(z-w)^2} +\frac{U_1 U_3'(w)+U_1' U_3(w)}{z-w}.
\eeaa
We also have
\beaa
V_2(z)U_4(w)&\sim & \frac{3 \alpha_0^4(N+1) N(N-1)(N-2)(N-3)}{2(z-w)^6}+\frac{\alpha_0^3(N+2)(N-1)(N-2)(N-3) U_1(w)}{(z-w)^5}\\
&&+\frac{\alpha_0^2(N+5)(N-2)(N-3) U_2(w)}{2(z-w)^4}+\frac{3 \alpha_0(N-3) U_3(w)}{(z-w)^3}+\frac{4 U_4(w)}{(z-w)^2}+\frac{U_4'(w)}{z-w}.
\eeaa
Note that we do not get the OPEs $V_1(z)U_4(w)$ and $V_2(z)U_4(w)$ from direct calculation, instead we get them from the results in \cite{pro1411}. Other OPEs are obtained from direct calculation here.
\section*{Appendix B: Schur functions and Jack polynomials}
Schur functions are defined in (\ref{hxi}) and (\ref{Slambda}). We list the first few of them:
 \beaa
&& S_{\Box}=p_1,\ S_{\begin{tikzpicture}
\draw [step=0.2](0,0) grid(.4,.2);
\end{tikzpicture}}=\frac{1}{2}p_1^2+\frac{1 }{2}p_{2},\ S_{\begin{tikzpicture}
\draw [step=0.2](0,0) grid(.2,.4);
\end{tikzpicture}}=\frac{1}{2}p_1^2- \frac{1 }{2}p_{2},\\
&&S_{\begin{tikzpicture}
\draw [step=0.2](0,0) grid(.6,.2);
\end{tikzpicture}}=\frac{1}{12}\left(2p_1^3+6p_1p_{2}+4p_{3}\right),\ S_{\begin{tikzpicture}
\draw [step=0.2](0,0) grid(.4,.2);
\draw [step=0.2](0,-0.2) grid(.2,0);
\end{tikzpicture}}=\frac{1}{6}\left(2p_1^3-2p_{3}\right),\\
&&S_{\begin{tikzpicture}
\draw [step=0.2](0,0) grid(.2,.6);
\end{tikzpicture}}=\frac{1}{12}\left(2p_1^3-6p_1p_{2}+4p_{3}\right),\\
&&S_{\begin{tikzpicture}
\draw [step=0.2](0,0) grid(.8,.2);
\end{tikzpicture}}=\frac{1}{24}(p_1^4+6p_1^2p_{2}+3p_{2}^2+8p_1p_{3}+6p_{4}),\\
&&S_{\begin{tikzpicture}
\draw [step=0.2](0,0) grid(.6,.2);
\draw [step=0.2](0,-0.2) grid(.2,0);
\end{tikzpicture}}= \frac{1}{8}(p_1^4+2p_1^2p_{2}-p_{2}^2-2p_{4}),\\
&&S_{\begin{tikzpicture}
\draw [step=0.2](0,0) grid(.4,.2);
\draw [step=0.2](0,-0.2) grid(.4,0);
\end{tikzpicture}}= \frac{1}{12}(p_1^4+3p_{2}^2-4p_1p_{3}),\\
&&S_{\begin{tikzpicture}
\draw [step=0.2](0,0) grid(.4,.2);
\draw [step=0.2](0,-0.2) grid(.2,0);
\draw [step=0.2](0,-0.4) grid(.2,-0.2);
\end{tikzpicture}}=\frac{1}{8}(p_1^4-2p_1^2p_{2}-p_{2}^2+2p_{4}),\\
&&S_{\begin{tikzpicture}
\draw [step=0.2](0,0) grid(.2,.8);
\end{tikzpicture}}= \frac{1}{24}(p_1^4-6p_1^2p_{2}+3p_{2}^2+8p_1p_{3}-6p_{4}).
\eeaa

The usual Jack polynomials $P_\lambda^\alpha,\ Q_\lambda^\alpha$ are defined in \cite{Mac}. To match the representation of the affine Yangian of $\mathfrak{gl}(1)$ on 3D Young diagrams, we defined the symmetric functions $Y_\lambda$ \cite{WBCW}. We list the first few of them:
\begin{eqnarray*}
Y_\Box&=&p_1,\\
 Y_{\begin{tikzpicture}
\draw [step=0.2](0,0) grid(.4,.2);
\end{tikzpicture}} &=&\frac{1}{h_1-h_2}p_{2}-\frac{h_2}{h_1-h_2}p_{1}^2,\\
Y_{\begin{tikzpicture}
\draw [step=0.2](0,0) grid(.2,.4);
\end{tikzpicture}}&=&\frac{1}{h_2-h_1}p_{2}-\frac{h_1}{h_2-h_1}p_{1}^2,\\
Y_{\begin{tikzpicture}
\draw [step=0.2](0,0) grid(.6,.2);
\end{tikzpicture}}&=&\frac{1}{2h_1-h_2}\frac{1}{h_1-h_2}(2p_{3}-3h_2p_{1}p_{2}+h_2^2p_{1}^3),\\
Y_{\begin{tikzpicture}
\draw [step=0.2](0,0) grid(.4,.2);
\draw [step=0.2](0,-0.2) grid(.2,0);
\end{tikzpicture}_{h_1h_2}}
&=&\frac{1}{h_2-2h_1}\frac{1}{h_1-h_2}(2p_{3}-2(h_1+h_2)p_{1}p_{2}+2h_1h_2p_{1}^3),\\
Y_{\begin{tikzpicture}
\draw [step=0.2](0,0) grid(.4,.2);
\draw [step=0.2](0,-0.2) grid(.2,0);
\end{tikzpicture}_{h_2h_1}}
&=&\frac{1}{h_1-2h_2}\frac{1}{h_2-h_1}(2p_{3}-2(h_1+h_2)p_{1}p_{2}+2h_1h_2p_{1}^3),\\
Y_{\begin{tikzpicture}
\draw [step=0.2](0,0) grid(.2,.6);
\end{tikzpicture}}&=&\frac{1}{2h_2-h_1}\frac{1}{h_2-h_1}(2p_{3}-3h_1p_{1}p_{2}+h_1^2p_{1}^3),\\
Y_{\begin{tikzpicture}
\draw [step=0.2](0,0) grid(.8,.2);
\end{tikzpicture}}
&=&\frac{1}{3h_1-h_2}\frac{1}{2h_1-h_2}\frac{1}{h_1-h_2}(6p_{4}-3h_2p_{2}^2-8h_2p_1p_{3}+6h_2^2p_1^2p_2-h_2^3p_{1}^4),\eeaa
\beaa
Y_{\begin{tikzpicture}
\draw [step=0.2](0,0) grid(.6,.2);
\draw [step=0.2](0,-0.2) grid(.2,0);
\end{tikzpicture}_{h_1,2h_1,h_2}}
&=&\frac{1}{3h_1-h_2}\frac{1}{2h_1-h_2}\frac{1}{h_1-h_2}(-6p_{4}+3h_2p_{2}^2+6(h_1+h_2)p_1p_{3}\nn\\
&&-(9h_1h_2+3h_2^2)p_1^2p_2+3h_1h_2^2p_{1}^4),\\
Y_{\begin{tikzpicture}
\draw [step=0.2](0,0) grid(.4,.2);
\draw [step=0.2](0,-0.2) grid(.2,0);
\draw [step=0.2](0,-0.4) grid(.2,-0.2);
\end{tikzpicture}_{h_2,2h_2,h_1}}
&=&\frac{1}{3h_2-h_1}\frac{1}{2h_2-h_1}\frac{1}{h_2-h_1}(-6p_{4}+3h_1p_{2}^2+6(h_1+h_2)p_1p_{3}\nn\\
&&-(9h_1h_2+3h_1^2)p_1^2p_2+3h_2h_1^2p_{1}^4),\eeaa
\beaa
Y_{\begin{tikzpicture}
\draw [step=0.2](0,0) grid(.4,.2);
\draw [step=0.2](0,-0.2) grid(.4,0);
\end{tikzpicture}_{h_1,h_2,h_1+h_2}}
&=&\frac{1}{h_2-2h_1}\frac{1}{(h_2-h_1)^3}(-2(h_1+h_2)p_{4}+2(h_1^2-h_1h_2+h_2^2)p_{2}^2+8h_1h_2p_1p_{3}\nn\\
&&-4h_1h_2(h_1+h_2)p_1^2p_2+2h_1^2h_2^2p_{1}^4),\\
Y_{\begin{tikzpicture}
\draw [step=0.2](0,0) grid(.2,.8);
\end{tikzpicture}}
&=&\frac{1}{3h_2-h_1}\frac{1}{2h_2-h_1}\frac{1}{h_2-h_1}(6p_{4}-3h_1p_{2}^2-8h_1p_1p_{3}+6h_1^2p_1^2p_2-h_1^3p_{1}^4).
\end{eqnarray*}
We still call $Y_\lambda$ Jack polynomials since they will become Jack polynomials when $h_1=\sqrt{\alpha},h_2=-1/\sqrt{\alpha}$. Note that the subscripts of the 2D Young diagrams show the expressions of $Y_\lambda$ depend on the box growth process. Treat 2D Young diagrams as special 3D Young diagrams which have one layer in $z$-axis, The notation $h_\Box=(x_\Box-1) h_2+ (y_\Box-1) h_1$ in subscript means adding a box in the position $(x_\Box, y_\Box)$. So for example, $\begin{tikzpicture}
\draw [step=0.2](0,0) grid(.4,.2);
\draw [step=0.2](0,-0.2) grid(.2,0);
\end{tikzpicture}_{h_1h_2}$ means this state is obtained from $\Box$ by
\[
\Box\rightarrow \begin{tikzpicture}
\draw [step=0.2](0,0) grid(.4,.2);
\end{tikzpicture}\rightarrow \begin{tikzpicture}
\draw [step=0.2](0,0) grid(.4,.2);
\draw [step=0.2](0,-0.2) grid(.2,0);
\end{tikzpicture}.
\]
 \section*{Data availability statement}
The data that support the findings of this study are available from the corresponding author upon reasonable request.

\section*{Declaration of interest statement}
The authors declare that we have no known competing financial interests or personal relationships that could have appeared to influence the work reported in this paper.

\section*{Acknowledgements}
This research is supported by the National Natural Science Foundation
of China under Grant No. 12101184, and supported by the Key Scientific Research Project in Colleges and Universities of Henan Province No. 22B110003. We would like to thank T. Prochazka for useful discuss about the calculations of OPEs. We also would like to thank P. Agarwal and S. Krivonos for useful discuss about how to use mathematica.

\end{document}